\documentclass[aps,pre,reprint,superscriptaddress,longbibliography]{revtex4-2}

\usepackage{amsmath,amssymb,amsthm,bm}
\usepackage{graphicx}
\usepackage{xcolor}
\usepackage[colorlinks=true,allcolors=blue]{hyperref}

\newtheorem{theorem}{Theorem}
\newtheorem{proposition}{Proposition}
\newtheorem{remark}{Remark}

\newcommand{\DKL}{D_{\mathrm{KL}}}
\newcommand{\Pfun}{P}
\newcommand{\Snull}{S_{\mathrm{null}}}
\newcommand{\Strue}{S_{\mathrm{true}}}
\newcommand{\Spop}{S_0}
\newcommand{\lcpop}{\lambda_c^{\mathrm{pop}}}
\newcommand{\Rres}{R}
\newcommand{\Dloc}{D_{\mathrm{KL,loc}}^{\min}}
\newcommand{\Dlocarma}{D_{\mathrm{KL,loc},(1,1)}^{\min}}

\begin{document}

\title{Timescale Coalescence Makes Hidden Persistent Forcing Spectrally Dark}

\author{Yuda Bi}
\email{ybi3@student.gsu.edu}
\affiliation{Translational Research in Neuroimaging and Data Science (TReNDS), Georgia State University, Atlanta, Georgia 30303, USA}
\author{Chenyu Zhang}
\affiliation{School of Medical Technology, Beijing Institute of Technology, Beijing 100081, China}
\author{Vince D. Calhoun}
\affiliation{Translational Research in Neuroimaging and Data Science (TReNDS), Georgia State University, Atlanta, Georgia 30303, USA}
\affiliation{School of Electrical and Computer Engineering, Georgia Institute of Technology, Atlanta, Georgia 30332, USA}

\date{\today}

\begin{abstract}
Under coarse observation, unresolved slow forcing can remain dynamically active yet locally invisible to reduced spectral inference. For a solvable driven AR$(1)$ benchmark, the local Whittle/Kullback--Leibler distance from the true spectrum to the best nearby one-pole surrogate obeys $\Dloc(\lambda)=C\lambda^4+O(\lambda^6)$, even though the observed spectrum itself is perturbed at $O(\lambda^2)$. The quartic onset is a geometric consequence of the reduced model manifold: the $O(\lambda^2)$ perturbation is partially absorbed by tangent-space reparametrization, and only the normal residual survives. We obtain $C$ in closed form for an AR$(1)$ hidden driver and show that $C$ vanishes as $(a-b)^2$ at timescale coalescence, identifying a spectrally \emph{dark} regime. We then show that this dark regime is not geometrically inevitable: for a non-degenerate AR$(2)$ hidden driver (second characteristic root $z_2\neq 0$), $C>0$ for all parameter values, including single-root coalescence, because the richer spectral structure cannot be absorbed by the two-dimensional tangent space. The quartic coefficient interpolates smoothly between the two cases as $C\sim z_2^4$ when the second characteristic root vanishes. Together, the AR$(1)$ and AR$(2)$ results yield a classification within the one-pole projection class: the quartic law and the boundary $\lcpop(N)\propto(\log N/N)^{1/4}$ are universal features of the projection geometry within this class, while the dark regime requires the hidden driver's spectrum to match the null family's pole structure.
\end{abstract}
\maketitle

\section{Introduction}
Reduced observations can hide slow physics rather than merely blur it. An unresolved slow mode may be dynamically active, inject low-frequency power, and yet still look indistinguishable from intrinsic persistence once the data are compressed to a coarse observable \cite{Hasselmann1976,FrankignoulHasselmann1977,PenlandSardeshmukh1995,MajdaTimofeyevVandenEijnden1999,ChorinHaldKupferman2000}. The same ambiguity appears in nonequilibrium inference under partial observation, where hidden slow variables can mask dissipation or irreversibility \cite{RoldanParrondo2010,MehlEtAl2012,SkinnerDunkel2021,Seifert2019}. This is not just an interpretive nuisance: it changes whether an apparently persistent reduced mode should be read as intrinsic memory or as a surrogate for unresolved forcing \cite{IsraeliGoldenfeld2006,MachtaEtAl2013}.

The ambiguity becomes especially sharp in the frequency domain, because excess low-frequency power is already a reduced statistic rather than a direct fingerprint of mechanism \cite{Priestley1981,GhilEtAl2002}. A broad low-frequency peak may reflect intrinsic autoregressive memory, persistent hidden forcing, or both at once. In reduced spectral inference one does not compare the observed data to a full hidden-variable truth model. Instead, one refits the spectrum to a simpler nearby surrogate, often a one-pole description, using Whittle likelihood or related criteria, and then judges alternatives through penalties such as AIC or BIC \cite{Whittle1953,Dzhaparidze1986,TaniguchiKakizawa2000,Akaike1974,Schwarz1978,HannanRissanen1982}. The relevant question is therefore not simply whether hidden forcing perturbs the spectrum, but whether that perturbation survives projection onto the best nearby reduced null.

This is a concrete instance of model-manifold compression under coarse graining: perturbation directions tangent to the reduced family are locally absorbed by reparametrization, whereas only the normal component remains asymptotically visible \cite{ChorinHaldKupferman2000,GivonKupfermanStuart2004,IsraeliGoldenfeld2006,MachtaEtAl2013}. A hidden driver can therefore generate a genuine $O(\lambda^2)$ spectral deformation while leaving no $O(\lambda^2)$ distinguishability after projection. This matters because the inference failure is systematic rather than noisy: more data do not help until the hidden perturbation develops a visible normal component. In that sense, the hidden signature becomes \emph{spectrally dark}: dynamically present, but locally invisible after coarse-grained refitting. The same tangent-absorption geometry is already suggested by driven Ornstein--Uhlenbeck precursors at timescale matching, but the discrete-time setting used here is the simplest one in which the projection algebra closes exactly on a compact frequency domain.

What has been missing is an exact solvable benchmark for when coarse-grained refitting absorbs a hidden slow perturbation. Existing latent-variable and partially observed inference frameworks can detect hidden structure numerically or asymptotically in specific settings \cite{Rabiner1989,EphraimMerhav2002,AllmanMatiasRhodes2009}, but they do not isolate in closed form when a hidden slow perturbation is absorbed by motion along the reduced model manifold. In particular, they do not determine whether detectability must begin at the same order as the raw spectral deformation itself.

Here we provide that benchmark for an analytically tractable driven AR$(1)$-by-AR$(1)$ model and then classify when the associated dark regime does and does not occur. We prove a local theorem in the one-pole projection class: the Whittle/Kullback--Leibler divergence from the true spectrum to the best nearby one-pole surrogate satisfies $D_{\mathrm{loc}}(\lambda)=C\lambda^4+O(\lambda^6)$, even though hidden forcing perturbs the observed spectrum at $O(\lambda^2)$. The exact coefficient $C$ vanishes as $(a-b)^2$ when the hidden and intrinsic timescales coalesce, yielding a leading population boundary of order $(\log N/N)^{1/4}$. We then show that the dark regime is structurally specific rather than geometrically inevitable: replacing the AR$(1)$ hidden driver by an AR$(2)$ driver preserves the quartic onset but eliminates the coalescence suppression entirely. The quartic coefficient remains strictly positive for any non-degenerate AR$(2)$ driver ($z_2\neq 0$), including at single-root coalescence, and interpolates smoothly to the AR$(1)$ limit as the second characteristic root vanishes. Together, these results identify the quartic law as a universal feature of the one-pole projection geometry within this class and the dark regime as a structural coincidence that requires the hidden driver's spectrum to have the same pole structure as the null family. Figures~\ref{fig:mechanism}--\ref{fig:ar2-sweep} show the mechanism, the exact quartic law, the associated operational crossover, and the AR$(2)$ classification.
\section{Solvable Model and Geometric Mechanism}

We study the discrete-time benchmark
\begin{equation}
X_{t+1}=aX_t+\lambda F_t+\epsilon_t,
\qquad
F_{t+1}=bF_t+\eta_t,
\label{eq:model}
\end{equation}
with $|a|<1$, $|b|<1$, $\epsilon_t\sim N(0,\sigma_\epsilon^2)$, $\eta_t\sim N(0,\sigma_\eta^2)$, and $\epsilon\perp\eta$. Only $X_t$ is observed. The null family is the one-pole spectrum
\begin{equation}
\Snull(\omega;\tilde a,\tilde \sigma^2)
=
\frac{\tilde \sigma^2}{\Pfun_{\tilde a}(\omega)},
\qquad
\Pfun_c(\omega)=1-2c\cos\omega+c^2,
\label{eq:null-spectrum}
\end{equation}
whereas the exact observed spectrum is
\begin{equation}
\Strue(\omega;\lambda)
\!=\!
\begin{aligned}
\frac{\sigma_\epsilon^2}{\Pfun_a(\omega)}
&+
\frac{\lambda^2\sigma_\eta^2}{\Pfun_a(\omega)\Pfun_b(\omega)} \\
&=
\Spop(\omega)\bigl(1+\lambda^2 h(\omega)\bigr),
\end{aligned}
\label{eq:true-spectrum}
\end{equation}
with $\Spop(\omega)=\sigma_\epsilon^2/\Pfun_a(\omega)$ and $h(\omega)=\sigma_\eta^2/[\sigma_\epsilon^2\Pfun_b(\omega)]$. The hidden driver therefore enters as an $O(\lambda^2)$ perturbation of the null spectrum. In numerical work we subtract the mean and drop the DC bin; the technical rationale is given in Appendix~C.

To turn Eq.~\eqref{eq:true-spectrum} into a detectability statement, one must compare the true spectrum not to the null point itself but to the best nearby one-pole fit. Write
\begin{equation}
\tilde a=a+\delta a,
\qquad
\tilde \sigma^2=\sigma_\epsilon^2+\delta \sigma^2.
\label{eq:local-params}
\end{equation}
Then the relative null spectrum expands as
\begin{equation}
\frac{\Snull(\omega;\tilde a,\tilde \sigma^2)}{\Spop(\omega)}
=
1+u\tilde e_1(\omega)+v\tilde e_2(\omega)+O(u^2+v^2+uv),
\label{eq:tangent-expansion}
\end{equation}
where
\begin{equation}
\begin{aligned}
u &= \frac{\delta \sigma^2}{\sigma_\epsilon^2}, \qquad
v = \delta a,\\
\tilde e_1(\omega) &= 1, \qquad
\tilde e_2(\omega) = \frac{2(\cos\omega-a)}{\Pfun_a(\omega)}.
\end{aligned}
\label{eq:tangent-basis}
\end{equation}
The one-pole manifold therefore has a two-dimensional tangent space $\mathcal{T}=\mathrm{span}\{\tilde e_1,\tilde e_2\}$ at the null point. In these coordinates the relevant population divergence is
\begin{equation}
\DKL(S_1\|S_2)
\!=\!
\frac{1}{4\pi}
\int_{-\pi}^{\pi}
\left[
\frac{S_1(\omega)}{S_2(\omega)}
-\log\frac{S_1(\omega)}{S_2(\omega)}-1
\right]d\omega.
\label{eq:DKL}
\end{equation}
where the normalized $L^2$ inner product is $\langle f,g\rangle_{L^2}:=\frac{1}{2\pi}\int_{-\pi}^{\pi}f(\omega)g(\omega)\,d\omega$.
Using the local scalar expansion $(1+\delta)-\log(1+\delta)-1=\delta^2/2+O(\delta^3)$, the fitting problem reduces to
\begin{equation}
\begin{aligned}
\DKL\!\left(
\Strue(\cdot;\lambda)\,\|\,\Snull(\cdot;\tilde a,\tilde \sigma^2)
\right)
&=
\frac{1}{4}
\bigl\|
\lambda^2 h-u\tilde e_1-v\tilde e_2
\bigr\|_{L^2}^2 \\
&\quad + O\!\left((|u|+|v|+\lambda^2)^3\right).
\end{aligned}
\label{eq:local-reduction}
\end{equation}
Equation~\eqref{eq:local-reduction} is the local geometry in one line: the best one-pole fit is the orthogonal projection of the leading perturbation onto the tangent space of the null manifold.
The same projection data also determine how the best one-pole surrogate moves along that manifold:
\begin{equation*}
\begin{aligned}
\tilde \sigma^{2*}(\lambda)
&=
\sigma_\epsilon^2
+
\lambda^2\frac{\sigma_\eta^2}{1-b^2}
+ O(\lambda^4),\\
\tilde a^*(\lambda)
&=
a
+
\lambda^2
\frac{\sigma_\eta^2}{\sigma_\epsilon^2}
\frac{b(1-a^2)}{(1-ab)(1-b^2)}
+ O(\lambda^4).
\end{aligned}
\end{equation*}
These pseudo-true shifts make tangent absorption concrete: the best one-pole fit drifts along the null manifold exactly in the direction predicted by the projection picture.

\begin{figure}[t]
  \centering
  \includegraphics[width=\columnwidth]{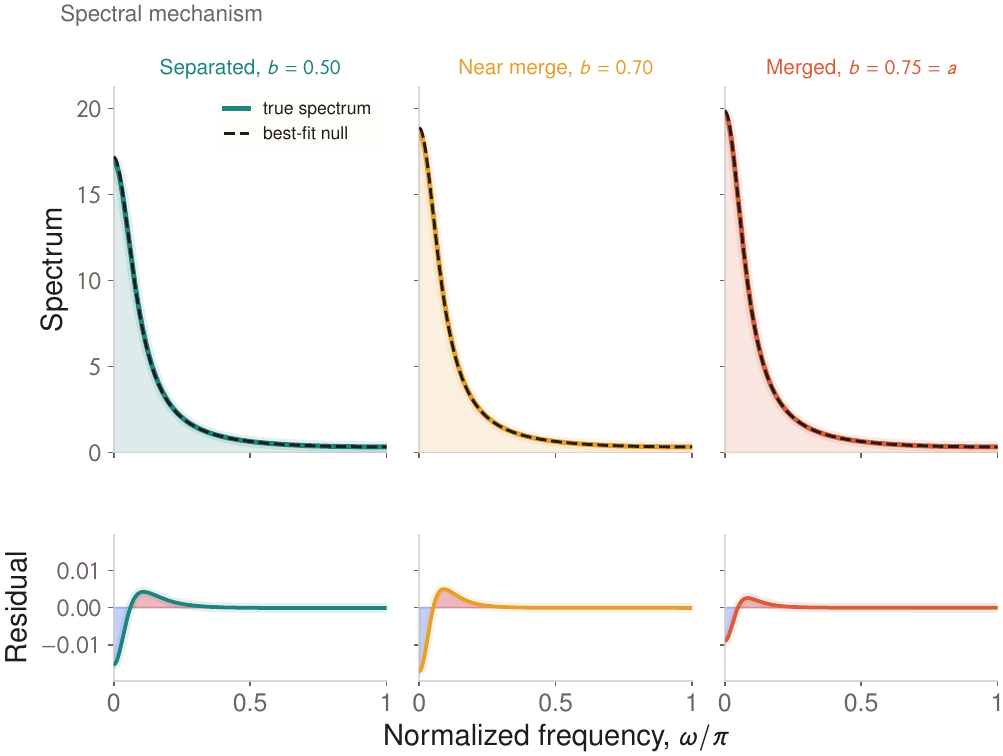}
  \caption{Pole coalescence hides the driver. Parameters: $a=0.75$, $\lambda=0.12$, $\sigma_\epsilon^2=\sigma_\eta^2=1$, with hidden persistence $b=0.50$, $0.70$, and $0.75=a$ from top to bottom. The exact driven spectrum is compared with the best-fit one-pole null spectrum for separated, near-coalescent, and merged timescales; the merged case corresponds to exact coalescence, $b=a$. The residual spectral shape is pronounced away from coalescence and is strongly suppressed as the hidden and intrinsic poles merge.}
  \label{fig:mechanism}
\end{figure}

Figure~\ref{fig:mechanism} shows why this is not yet the detectability law. Away from timescale matching, the best one-pole fit leaves a visible residual spectral shape. At pole coalescence, by contrast, the residual is strongly suppressed because the leading perturbation becomes tangent to the one-pole manifold and is reabsorbed by the best null reparametrization.

\section{Quartic Detectability Law}

The tangent basis in Eq.~\eqref{eq:tangent-basis} is orthogonal in the normalized $L^2$ inner product,
\begin{equation}
\langle \tilde e_1,\tilde e_2\rangle=0,
\qquad
\|\tilde e_1\|^2=1,
\qquad
\|\tilde e_2\|^2=\frac{2}{1-a^2},
\label{eq:gram-data}
\end{equation}
so the local Hessian of the Whittle objective is positive definite. Writing
\begin{equation}
\Rres=h-\Pi_{\mathcal{T}}h
\label{eq:R-def}
\end{equation}
for the normal residual after tangent absorption, the nearby pseudo-true one-pole branch satisfies
\begin{equation}
\begin{aligned}
\Dloc(\lambda)
&=
\DKL\!\left(\Strue(\cdot;\lambda)\,\|\,\Snull(\cdot;\tilde a,\tilde \sigma^2)\right)
\Big|_{(\tilde a,\tilde \sigma^2)=(\tilde a^*(\lambda),\tilde \sigma^{2*}(\lambda))} \\
&=
\frac{\lambda^4}{4}\|\Rres\|_{L^2}^2+O(\lambda^6).
\end{aligned}
\label{eq:projection-law}
\end{equation}
Equation~\eqref{eq:projection-law} is the geometric content of the theorem: the $O(\lambda^2)$ perturbation survives only through its normal component, so detectability begins at quartic order. Because the tangent basis is orthogonal and the local Hessian is positive definite, the reduced Whittle objective admits a unique nearby minimizer branch for sufficiently small $|\lambda|$. The theorem therefore controls a well-defined pseudo-true one-pole continuation of the null model rather than an informal best-fit heuristic; the derivation of the tangent basis, projection coefficients, local minimizer, and DC-bin implementation is given in Appendices~C--E.

\begin{theorem}
\label{thm:main}
For each fixed $(a,b)$ with $|a|<1$ and $|b|<1$, there exists $\lambda_0(a,b)>0$ such that for $|\lambda|<\lambda_0(a,b)$ the local Whittle minimizer branch within the one-pole projection class exists uniquely and the associated local minimum satisfies
\begin{equation}
\Dloc(\lambda)
=
C(a,b,\sigma_\epsilon,\sigma_\eta)\lambda^4+O(\lambda^6),
\label{eq:main-theorem}
\end{equation}
with $C\ge 0$ and
\begin{equation}
C=0 \iff (a=b)\ \text{or}\ (b=0).
\label{eq:main-zero}
\end{equation}
\end{theorem}

For the solvable model in Eq.~\eqref{eq:model}, the projection integrals can be evaluated exactly in Appendix~D, so Eq.~\eqref{eq:projection-law} closes as
\begin{equation}
\Dloc(\lambda)
=
C(a,b,\sigma_\epsilon,\sigma_\eta)\lambda^4+O(\lambda^6),
\label{eq:quartic-law}
\end{equation}
with
\begin{equation}
C(a,b,\sigma_\epsilon,\sigma_\eta)
=
\frac{\sigma_\eta^4}{2\sigma_\epsilon^4}
\frac{b^2(a-b)^2}{(1-b^2)^3(1-ab)^2}.
\label{eq:quartic-coefficient}
\end{equation}
The zero set confirms Eq.~\eqref{eq:main-zero}: $C=0$ if and only if the hidden driver is white ($b=0$, the trivial case in which added flat power is fully absorbed by $\tilde\sigma^2$ reparametrization) or if the hidden and intrinsic poles coalesce ($a=b$). For persistent hidden drivers $(b\neq 0)$, the quartic coefficient is strictly positive away from coalescence and vanishes exactly on the timescale-matching line $a=b$. In the present solvable class,
\begin{equation}
C(a,b,\sigma_\epsilon,\sigma_\eta)\propto (a-b)^2
\qquad
\text{as}\quad b\to a.
\label{eq:coalescence-law}
\end{equation}
Equation~\eqref{eq:quartic-coefficient} is the main theorem-level statement. It proves a local quartic detectability law and fixes the exact prefactor for the nearby pseudo-true one-pole branch. Hidden forcing becomes hard to detect not because its amplitude disappears, but because the leading spectral deformation is absorbed by a tangent reparametrization of the reduced manifold. The factor $(a-b)^2$ is the signature of timescale coalescence: when the hidden pole merges with the intrinsic one, the normal component vanishes and the quartic coefficient is suppressed to zero. At exact coalescence, the quartic law degenerates and higher-order terms take over. The resulting gap between the $O(\lambda^2)$ dynamical perturbation and the $O(\lambda^4)$ distinguishability is the paper's central physical point: under coarse graining, hidden forcing can remain dynamically active yet spectrally dark.

\begin{figure}[t]
  \centering
  \includegraphics[width=\columnwidth]{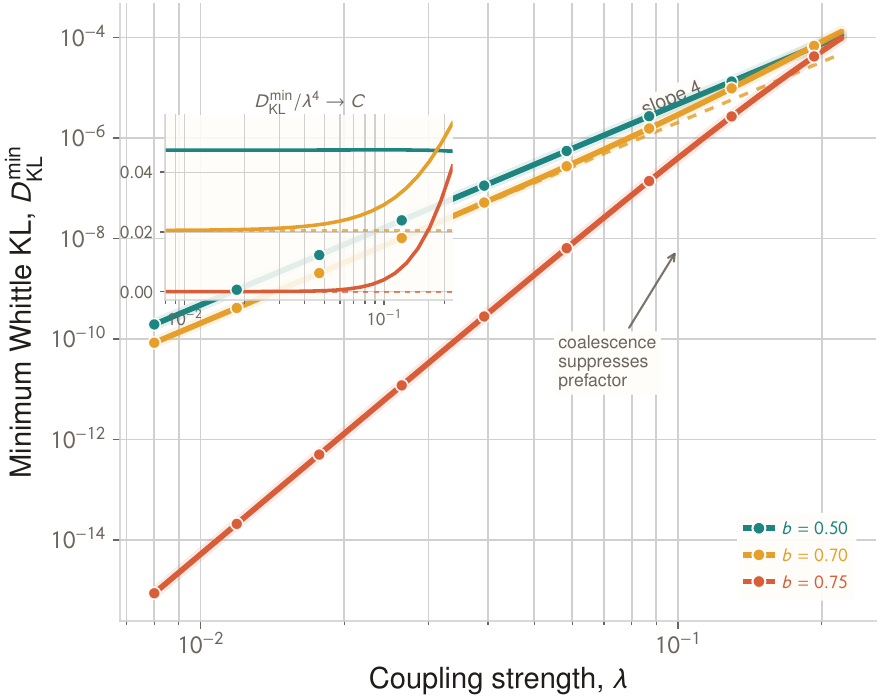}
  \caption{Detectability is quartic in coupling.
  Population-level numerical minimization of the one-pole
  projection problem for $a=0.75$,
  $\sigma_\epsilon^2=\sigma_\eta^2=1$, and $b=0.50$, $0.70$,
  $0.75=a$. Away from coalescence the quartic
  residual is visible; timescale matching suppresses
  it through a vanishing prefactor~$C$ (inset).}
  \label{fig:quartic}
\end{figure}

Figure~\ref{fig:quartic} displays the theorem in population numerics. All noncoalesced cases exhibit quartic onset, while the entire curve is pushed downward as $a\to b$ because the quartic prefactor is suppressed.
This is a local weak-coupling theorem for the nearby pseudo-true one-pole branch. The theorem does not prove that this local branch coincides with the global infimum over the full stationary one-pole class; that stronger statement is verified numerically in Appendix~I by exact global population minimization, which confirms that the tested parameter regimes remain on the local branch.

\emph{Mechanistic lesson.} The solvable benchmark suggests a broader local picture rather than a second theorem. A good reduced one-pole fit need not exclude hidden slow forcing. It may instead indicate that the leading hidden perturbation lies predominantly in a tangent direction of the reduced manifold and is locally absorbed by reparametrization \cite{ChorinHaldKupferman2000,GivonKupfermanStuart2004,IsraeliGoldenfeld2006}. Formally, if a reduced spectral family forms a smooth manifold through $S_0$, the true spectrum satisfies $S_{\mathrm{true}}(\lambda)=S_0(1+\lambda^2 h+O(\lambda^4))$, and the local Whittle Hessian is positive definite on the tangent space $\mathcal{T}$, then the same quadratic reduction yields
\begin{equation*}
D_{\mathrm{KL}}^{\min}(\lambda)
=
\frac{\lambda^4}{4}\|(I-\Pi_{\mathcal T})h\|^2+O(\lambda^6).
\end{equation*}
This formal criterion (proved as a proposition with explicit sufficient conditions in Appendix~F) is not a universal theorem for arbitrary model classes; the present solvable benchmark is its exact one-pole realization.

\section{Numerical and Operational Tests}

Equation~\eqref{eq:quartic-law} immediately yields a leading-order population model-selection boundary. Define $\lcpop(N)$ by balancing the local asymptotic KL gain against the BIC penalty,
\begin{equation}
N\,\Dloc(\lcpop)=\frac{\Delta k}{2}\log N.
\label{eq:bic-balance}
\end{equation}
For the effective two-pole alternative used in the Whittle-BIC tests, $\Delta k=2$, so Eqs.~\eqref{eq:quartic-law} and \eqref{eq:quartic-coefficient} give
\begin{equation}
\lcpop(N)
\!=\!
\left[
\frac{2\,\sigma_\epsilon^4(1-b^2)^3(1-ab)^2}
{\sigma_\eta^4 b^2(a-b)^2}
\frac{\log N}{N}
\right]^{1/4}
\left(1+o(1)\right).
\label{eq:lambda-c}
\end{equation}
Thus
\begin{equation}
\lcpop(N)\propto (\log N/N)^{1/4},
\qquad
\lcpop\propto |a-b|^{-1/2}.
\label{eq:boundary-scaling}
\end{equation}
The first scaling states that hidden forcing becomes detectable only at quartic order in coupling; the second states that detectability is parametrically suppressed as the hidden and intrinsic timescales merge.
Equations~\eqref{eq:bic-balance}--\eqref{eq:boundary-scaling} define a leading-order population boundary, not an exact finite-sample $50\%$-power threshold.
At fixed scaled coupling, Eq.~\eqref{eq:boundary-scaling} also implies $N/\log N\propto |a-b|^{-2}$, so the data cost of detection diverges quadratically as the hidden timescale approaches the intrinsic one. For equal noise variances, the boundary at $(a,b)=(0.90,0.80)$ gives $\lcpop(10^3)\approx 0.30$, while the near-coalescent choice $(0.90,0.88)$ raises it to $\lcpop(10^3)\approx 0.39$: shrinking the pole separation from $0.10$ to $0.02$ raises the leading-order sample-size requirement by a factor of $\approx 25$. The physical origin of this suppression is that the two tangent directions $\tilde e_1$ (amplitude) and $\tilde e_2$ (pole shift) can jointly mimic the spectral shape $\rho/\Pfun_b$ increasingly well as $b\to a$, until at exact coalescence the mimicry is perfect and the hidden driver becomes locally indistinguishable from a reparametrized null.

Three independent layers support Eqs.~\eqref{eq:quartic-law}--\eqref{eq:boundary-scaling}. First, exact global minimization over the full stationary one-pole family reproduces quartic onset, boundary scaling, and coalescence suppression, showing that the tested regimes remain on the same branch as the local theorem. Second, independent computer algebra (SymPy) reproduces the contour-integral identities, projection coefficients, and exact residual norm, fixing the theorem object without relying on a single derivation.

Third, the operational test. We simulate the full hidden-driver model in Eq.~\eqref{eq:model} at each point on a $(N,\lambda/\lcpop)$ grid, compute the periodogram (mean-subtracted, DC-dropped), and fit both the one-pole null $\Snull=\tilde\sigma^2/\Pfun_{\tilde a}$ and the two-pole alternative
\begin{equation}
S_{\mathrm{alt}}(\omega)
=
\frac{\sigma^2}{\Pfun_{\tilde a}(\omega)}
+
\frac{q}{\Pfun_{\tilde a}(\omega)\Pfun_{\tilde b}(\omega)}
\label{eq:alt-spectrum}
\end{equation}
by Whittle likelihood, with the hidden driver declared detected when $\mathrm{BIC}_{\mathrm{null}}-\mathrm{BIC}_{\mathrm{alt}}>0$ and $\Delta k=2$. All reported intervals are Wilson $95\%$ intervals for binomial detection frequencies over $200$ independent replications per grid point.

\begin{figure}[t]
  \centering
  \includegraphics[width=\columnwidth]{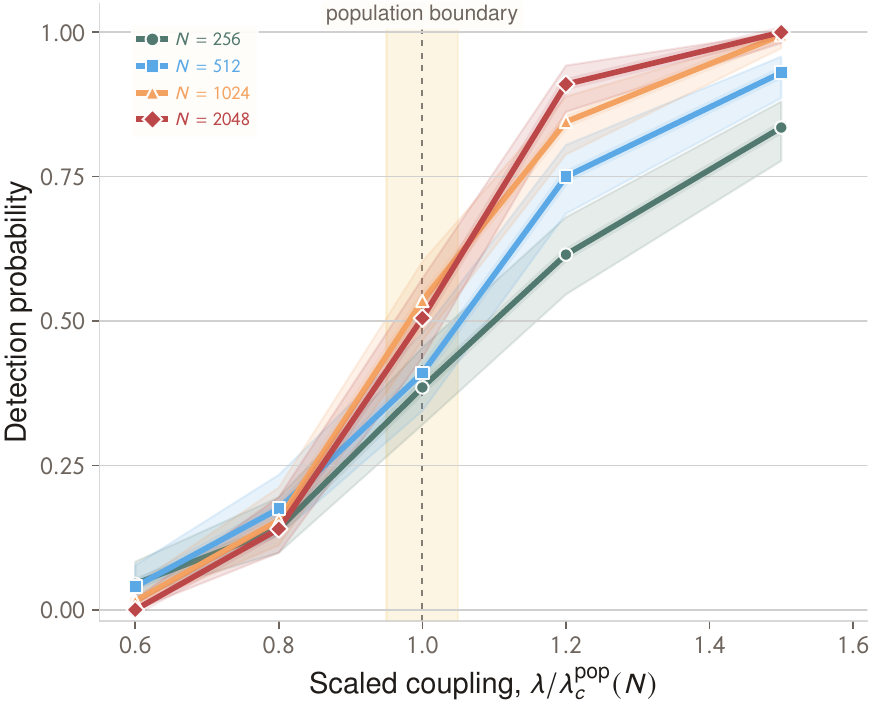}
  \caption{Model selection turns over near the predicted scaled boundary.
  Whittle-BIC detection probability is plotted against
  $\lambda/\lcpop(N)$ for the baseline parameters
  $(a,b,\sigma_\epsilon^2,\sigma_\eta^2)=(0.95,0.8,1,1)$
  and $N=256$--$2048$ ($200$ repetitions per point).
  The common crossover near unity is consistent with
  the asymptotic boundary but not with exact
  finite-size collapse.}
  \label{fig:operational}
\end{figure}

Figure~\ref{fig:operational} shows that Whittle-BIC model comparison turns over near the scaled boundary predicted by Eq.~\eqref{eq:lambda-c}. At $\lambda/\lcpop=1$, the detection probabilities are $0.385$, $0.410$, $0.535$, $0.505$ for $N=256$, $512$, $1024$, $2048$, with Wilson $95\%$ intervals of width $\approx 0.136$. The interpolated $50\%$-power locations are $r_{50}(N)=1.10$, $1.05$, $0.98$, $1.00$, consistent with a finite-$N$ drift toward the asymptotic boundary rather than a failure of the scaling law. Under the null ($\lambda=0$), the false-positive rate remains at or below $0.5\%$ for all tested $N$, confirming that the BIC criterion is well calibrated. Additional stress tests across parameter regimes, mild Student-$t$ misspecification, and the enriched ARMA$(1,1)$ null are reported in Appendix~J; they support the same interpretation without changing the main-text claim.

\section{Universality and Specificity of the Dark Regime}

The quartic detectability law in Eq.~\eqref{eq:quartic-law} was derived for a specific hidden driver, the AR$(1)$ process with coefficient $b$. Two independent extensions reveal which features of the result are universal within the projection class and which depend on the driver structure.

\subsection{Enriched null}

When the null family is enlarged from AR$(1)$ to ARMA$(1,1)$ (one pole plus one zero), the quartic onset survives with a reduced coefficient $C_{(1,1)}=b^2 C$ (Appendix~G). Enlarging the null absorbs one additional tangent component of $h$ but does not eliminate the mechanism.

\subsection{AR$(2)$ hidden driver}
\label{sec:ar2}

A more informative extension replaces the hidden driver itself. Consider the AR$(2)$ model
\begin{equation}
X_{t+1}=aX_t+\lambda F_t+\epsilon_t,
\qquad
F_{t+1}=b_1 F_t+b_2 F_{t-1}+\eta_t,
\label{eq:ar2-model}
\end{equation}
with characteristic roots $z_1,z_2$ satisfying $z^2-b_1 z-b_2=0$. The observed spectrum is $\Strue(\omega;\lambda)=\sigma_\epsilon^2/\Pfun_a(\omega)+\lambda^2\sigma_\eta^2/[\Pfun_a(\omega)Q_b(\omega)]$, with $Q_b(\omega)=|1-b_1 e^{-i\omega}-b_2 e^{-2i\omega}|^2$, so the perturbation function becomes $h(\omega)=\sigma_\eta^2/[\sigma_\epsilon^2 Q_b(\omega)]$. Since the null manifold remains one-pole AR$(1)$, the tangent space and the quartic-law identity $\Dloc(\lambda)=\frac{\lambda^4}{4}\|\Rres\|^2+O(\lambda^6)$ are unchanged; only the inner products involving $h$ differ. In particular, the threshold scaling $\lcpop(N)\propto(\log N/N)^{1/4}$ is preserved.

The key new result concerns coalescence. Writing $\rho=\sigma_\eta^2/\sigma_\epsilon^2$ for the noise ratio, the AR$(1)$ dark regime exists because the perturbation $h=\rho/\Pfun_b$ at $b=a$ reduces to $h=\rho/\Pfun_a$, which is exactly a linear combination of the tangent directions $\tilde e_1$ and $\tilde e_2$. The tangent space therefore absorbs $h$ completely and $\Rres=0$.

For the AR$(2)$ driver, even when one root satisfies $z_1=a$, the perturbation
\begin{equation}
h(\omega)=\frac{\rho}{\Pfun_a(\omega)\Pfun_{z_2}(\omega)}
\label{eq:ar2-h-coal}
\end{equation}
still retains the factor $1/\Pfun_{z_2}(\omega)$. This extra factor introduces frequency dependence outside
\[
\mathcal{T}=\mathrm{span}\{\tilde e_1,\tilde e_2\}.
\]
Since the tangent space is two-dimensional whereas $h$ in Eq.~\eqref{eq:ar2-h-coal} carries three independent spectral-shape degrees of freedom, the identity
\[
h=\alpha \tilde e_1+\beta \tilde e_2
\]
is generically overdetermined and therefore has no solution when $z_2\neq 0$. It follows that the quartic coefficient remains strictly positive:
\begin{equation}
C_{\mathrm{AR}(2)}>0.
\label{eq:ar2-no-dark}
\end{equation}
Hence no spectrally dark regime exists for any nonwhite stationary AR$(2)$ driver with $z_2\neq 0$. This conclusion holds uniformly over the full parameter space, for all values of $a$, and in both the overdamped (real-root) and oscillatory (complex-conjugate-root) regimes.

Population-level minimization of the Whittle KL for the AR$(2)$-driven model over the coupling range $\lambda\in[0.005,0.15]$ confirms the quartic onset with log-log slopes of $3.87$ (overdamped, roots $0.70$ and $0.50$), $3.95$ (oscillatory, $|r|\approx 0.71$), and $3.81$ (persistent oscillatory, $|r|\approx 0.77$). The deviations from $4.0$ are consistent with $O(\lambda^6)$ corrections at this coupling range. At exact single-root coalescence ($z_1=a=0.95$, $z_2=0.5$), the quartic coefficient is $C_{\mathrm{AR}(2)}=0.975$, demonstrating that the dark regime is absent. The threshold scaling $\lcpop(N)\propto(\log N/N)^{1/4}$ is confirmed with a fitted exponent of $-0.251$. Whittle-BIC Monte Carlo experiments using the misspecified two-pole alternative in Eq.~\eqref{eq:alt-spectrum} show detection-probability turnover near the predicted boundary, confirming that the operational interpretation carries over from the AR$(1)$ baseline.

\begin{figure}[t]
  \centering
  \includegraphics[width=\columnwidth]{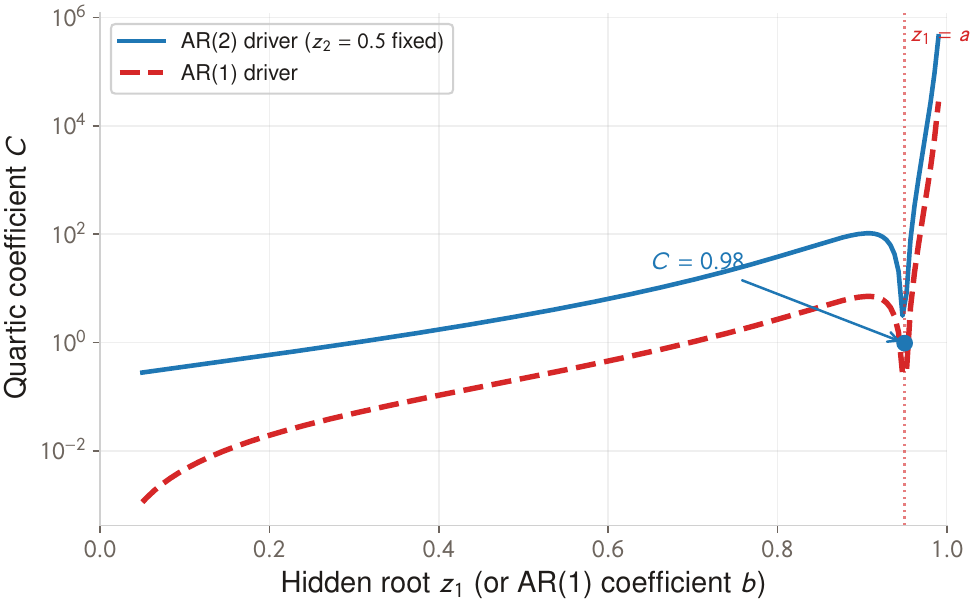}
  \caption{No coalescence suppression for the AR$(2)$ driver.
  The quartic coefficient $C_{\mathrm{AR}(2)}$ is plotted as a function of the first characteristic root $z_1$, with $z_2=0.5$ fixed and $a=0.95$.
  Unlike the AR$(1)$ case, $C$ remains strictly positive at $z_1=a$ (dashed line).
  For the AR$(1)$ driver, $C=0$ at $b=a$ because $h=\rho/\Pfun_a\in\mathcal{T}$; for the AR$(2)$ driver, the extra factor $1/\Pfun_{z_2}$ prevents full tangent absorption.}
  \label{fig:ar2-sweep}
\end{figure}

Figure~\ref{fig:ar2-sweep} shows the quartic coefficient as a function of $z_1$ with $z_2=0.5$ fixed: $C$ remains strictly positive everywhere, including at $z_1=a$.

\subsection{Smooth interpolation and classification}

The degenerate limit $z_2=0$ reduces the AR$(2)$ driver to AR$(1)$ with $b=z_1$. In this limit $\Pfun_{z_2}(\omega)\to\Pfun_0(\omega)=1$, so $h\to\rho/\Pfun_{z_1}$ and the AR$(1)$ coalescence is recovered. Population-level numerics confirm that the quartic coefficient interpolates smoothly:
\begin{equation}
C_{\mathrm{AR}(2)}(z_1\!=\!a,z_2)\;\sim\; C_0\,z_2^4
\qquad\text{as }z_2\to 0,
\label{eq:z2-scaling}
\end{equation}
providing a continuous bridge from the AR$(1)$ dark limit $C=0$ to the AR$(2)$ immunity $C>0$. Table~\ref{tab:z2-interpolation} illustrates this interpolation numerically. The quartic scaling in $z_2$ reflects the fact that the first two orders of the perturbation away from $z_2=0$ are absorbed by the tangent space, and only the quadratic residual squared (i.e., $O(z_2^4)$ in $\|\Rres\|^2$) survives.

\begin{table}[t]
\caption{Smooth interpolation of the quartic coefficient from the AR$(1)$ dark limit to the AR$(2)$ regime, at $z_1=a=0.95$ and $\sigma_\epsilon^2=\sigma_\eta^2=1$. The ratio $C/z_2^4$ stabilizes at small $z_2$, confirming the asymptotic quartic scaling in Eq.~\eqref{eq:z2-scaling}; deviations at larger $z_2$ reflect higher-order terms in the expansion.}
\label{tab:z2-interpolation}
\begin{ruledtabular}
\begin{tabular}{ccc}
$z_2$ & $C_{\mathrm{AR}(2)}$ & $C/z_2^4$ \\
\hline
$0$ (AR(1) limit) & $\approx 0$ & --- \\
$0.01$ & $5.2\times 10^{-9}$ & $0.52$ \\
$0.05$ & $3.8\times 10^{-6}$ & $0.61$ \\
$0.10$ & $7.7\times 10^{-5}$ & $0.77$ \\
$0.30$ & $2.1\times 10^{-2}$ & $2.54$ \\
$0.50$ & $9.8\times 10^{-1}$ & $15.6$ \\
\end{tabular}
\end{ruledtabular}
\end{table}

These results yield a complete classification within the one-pole projection class:
\begin{enumerate}
\item \emph{Quartic onset} $\Dloc=C\lambda^4+O(\lambda^6)$ is universal within the one-pole projection class: it depends only on the null manifold geometry, not on the hidden dynamics.
\item \emph{Spectrally dark regime} $C=0$ requires the perturbation $h$ to lie entirely in $\mathcal{T}$. Among the AR$(1)$ and AR$(2)$ hidden drivers studied here, this occurs only for the AR$(1)$ driver at $b=a$, where $h=\rho/\Pfun_a$ is exactly a linear combination of $\tilde e_1$ and $\tilde e_2$.
\item \emph{Richer hidden dynamics} ($z_2\neq 0$) are always detectable at quartic order, with a smooth $O(z_2^4)$ onset of detectability as the second root departs from zero.
\end{enumerate}
Numerical and Monte Carlo validation for the AR$(2)$ case, including threshold scaling and Whittle-BIC power curves, is reported in Appendix~\ref{app:ar2}.

\section{Discussion}

The distinction between what is universal within the one-pole projection class and what is specific to the driver structure is the main structural insight. A good reduced one-pole fit need not exclude hidden slow forcing, but the reason for under-detection depends on the driver complexity. For the simplest single-pole drivers, the hidden perturbation can be fully tangent to the null manifold and thus locally invisible regardless of sample size. For richer multi-pole drivers, the perturbation always has a nonzero normal component, so the hidden forcing is in principle detectable at quartic order given sufficient data. The classification is therefore not merely an abstract decomposition: it determines whether increasing sample size can eventually resolve the hidden driver or whether the dark regime poses a fundamental barrier.

For practical inference, this classification carries a specific message. In many physical systems the unresolved slow forcing is unlikely to have a purely single-pole spectral shape: ocean heat transport, mesoscale eddies, and hidden molecular modes often exhibit richer spectral structures than a single AR$(1)$ process. The dark regime is then a codimension-one phenomenon---it requires exact spectral shape matching between hidden driver and null family---and any departure from single-pole structure restores detectability at quartic order. In that sense, the dark regime identified in the AR$(1)$ benchmark is a worst case rather than the generic situation. The quartic onset, by contrast, is the robust feature within this projection class: it sets the fundamental rate at which hidden slow forcing becomes visible as coupling increases, regardless of the driver's internal complexity.

To illustrate the order of magnitude, consider a climate-style setting with monthly data over $50$ years ($N=600$) and intrinsic persistence $a=0.90$. For a well-separated hidden driver with $b=0.80$ and unit noise variances, the population boundary is $\lcpop(600)\approx 0.30$. A hidden forcing coupled at $\lambda=0.25$ is therefore below the quartic boundary and would be missed by a standard one-pole BIC test with this sample length. Doubling the record to $N=1200$ only reduces the boundary to $\lcpop(1200)\approx 0.25$, reflecting the slow $(\log N/N)^{1/4}$ scaling. These are not application-specific claims but order-of-magnitude consequences of the exact boundary law.

The claim is deliberately local. It concerns weak coupling around $\lambda=0$ for Gaussian, stationary hidden drivers and the one-pole, or one-pole ARMA$(1,1)$, reduced classes. It does not imply an onset theorem for arbitrary latent models, because in other settings the leading visible order can differ. A complementary question is whether retaining cross-spectral information between multiple observed channels can break the dark regime even for single-pole drivers. Within those bounds, the benchmark isolates the mechanism cleanly and provides a concrete starting point for richer model-manifold studies.

\bibliography{prl_refs}

\clearpage
\onecolumngrid
\appendix
\setcounter{equation}{0}
\setcounter{figure}{0}
\setcounter{table}{0}
\renewcommand{\theequation}{S\arabic{equation}}
\renewcommand{\thefigure}{S\arabic{figure}}
\renewcommand{\thetable}{S\arabic{table}}

\begin{center}
{\Large\bfseries Appendices}\\[0.35em]
{\large\bfseries Timescale Coalescence Makes Hidden Persistent Forcing Spectrally Dark}
\end{center}

\section*{Appendix Contents}

\begin{itemize}
  \item Appendix A: Supplementary Overview and Literature Positioning
  \item Appendix B: Exact Spectrum and Relative Perturbation
  \item Appendix C: Local Whittle Geometry of the One-Pole Manifold
  \item Appendix D: Projection Coefficients and Pseudo-True Shifts
  \item Appendix E: Proof of the Quartic Detectability Theorem
  \item Appendix F: Corollaries and the Local Tangent-Absorption Criterion
  \item Appendix G: Enriched Null Families and the ARMA$(1,1)$ Corollary
  \item Appendix H: Independent Symbolic Verification
  \item Appendix I: Exact Numerical KL Validation
  \item Appendix J: Operational Monte Carlo Validation
  \item Appendix K: Continuous-Time Foundation and Scope
  \item Appendix L: Extension to an AR$(2)$ Hidden Driver
\end{itemize}

\section{Supplementary Overview and Literature Positioning}

These appendices record the derivations supporting the main-text theorem and its numerical checks. The physical question is whether an observed slow mode is intrinsically persistent or only appears persistent because it is driven by an unresolved persistent force. That ambiguity is central to stochastic climate modeling \cite{Hasselmann1976,FrankignoulHasselmann1977,PenlandSardeshmukh1995}, reduced nonequilibrium inference with hidden slow variables \cite{RoldanParrondo2010,MehlEtAl2012,SkinnerDunkel2021,Seifert2019}, and thermodynamic inference under coarse observation \cite{KawaiParrondoVandenBroeck2007,EspositoVandenBroeck2010a,HartichBaratoSeifert2014,BaratoSeifert2015}.

Energy-balance climate models, stochastic mode-reduction schemes, and advanced spectral analyses reinforce the same point from complementary directions \cite{NorthCahalanCoakley1981,MajdaTimofeyevVandenEijnden1999,GhilEtAl2002}. In nonequilibrium settings, fluctuation theorems, thermodynamics of information, and marginal-observer constructions provide parallel motivation for asking what reduced observations can still distinguish \cite{Crooks1999,Jarzynski1997,Seifert2012,Jarzynski2011,ParrondoHorowitzSagawa2015,HorowitzEsposito2014,ShiraishiSagawa2015,PolettiniEsposito2017,BiskerEtAl2017,Schnakenberg1976}.

The technical setting is the intersection of Whittle spectral inference \cite{Whittle1953,TaniguchiKakizawa2000}, classical spectral-analysis practice \cite{Priestley1981,BrockwellDavis2016,ShumwayStoffer2017}, information-criterion model selection \cite{Akaike1974,Schwarz1978,BurnhamAnderson2002,ClaeskensHjort2008,StoicaSelen2004,HannanRissanen1982,HurvichTsai1989,Dzhaparidze1986}, hidden-cause inference in time series \cite{Granger1969,Sims1972,Geweke1982}, and latent-state recovery \cite{Rabiner1989,Leroux1992,CappeMoulinesRyden2005,EphraimMerhav2002,HsuKakadeZhang2012,AllmanMatiasRhodes2009,VerbekeMolenberghs2017}. What is specific to the present work is the exact quartic detectability law and the explicit coalescence mechanism within the one-pole projection class.

What is proved exactly in the appendices below is:
\begin{itemize}
  \item a local quartic detectability law in the one-pole projection class;
  \item a closed form for the quartic coefficient;
  \item a quantitative coalescence law $C\propto(a-b)^2$;
  \item an enriched-null ARMA$(1,1)$ corollary that makes model-class dependence explicit;
  \item a leading-order population boundary $\lcpop\propto (\log N/N)^{1/4}$;
  \item independent symbolic, numerical, and operational validation of these statements.
\end{itemize}
What is not claimed is:
\begin{itemize}
  \item a universal theorem for arbitrary nonlinear or non-Gaussian hidden-variable architectures;
  \item a proof of the first nonzero post-quartic term exactly at strict coalescence;
  \item a minimax impossibility theorem for all possible tests;
  \item a fully explicit bridge to every application-specific parametrization.
\end{itemize}

\section{Exact Spectrum and Relative Perturbation}

The model in Eq.~\eqref{eq:model} can be written with the lag operator $L$ as
\begin{equation}
(1-aL)X_t=\lambda F_{t-1}+\epsilon_t,
\qquad
(1-bL)F_t=\eta_t.
\label{eqS:lag-form}
\end{equation}
In frequency space,
\begin{equation}
X(\omega)
\!=\!
\frac{\epsilon(\omega)}{1-ae^{-i\omega}}
+
\lambda\,
\frac{e^{-i\omega}\eta(\omega)}
{(1-ae^{-i\omega})(1-be^{-i\omega})}.
\label{eqS:freq-form}
\end{equation}
Because $\epsilon$ and $\eta$ are independent and white, the observed spectrum is
\begin{equation}
\Strue(\omega;\lambda)
\!=\!
\begin{aligned}
\frac{\sigma_\epsilon^2}{|1-ae^{-i\omega}|^2}
&+
\frac{\lambda^2\sigma_\eta^2}
{|1-ae^{-i\omega}|^2|1-be^{-i\omega}|^2} \\
&=
\frac{\sigma_\epsilon^2}{\Pfun_a(\omega)}
+
\frac{\lambda^2\sigma_\eta^2}{\Pfun_a(\omega)\Pfun_b(\omega)}.
\end{aligned}
\label{eqS:true-spectrum}
\end{equation}
Writing $\Spop(\omega)=\sigma_\epsilon^2/\Pfun_a(\omega)$ gives the relative perturbation form
\begin{equation}
\Strue(\omega;\lambda)
=
\Spop(\omega)\bigl(1+\lambda^2 h(\omega)\bigr),
\qquad
h(\omega)=\frac{\sigma_\eta^2}{\sigma_\epsilon^2}\frac{1}{\Pfun_b(\omega)}.
\label{eqS:relative}
\end{equation}

\section{Local Whittle Geometry of the One-Pole Manifold}

\subsection{Relative coordinates near the null point}

Let
\begin{equation}
\tilde a=a+\delta a,
\qquad
\tilde \sigma^2=\sigma_\epsilon^2+\delta \sigma^2.
\label{eqS:local-params}
\end{equation}
Then
\begin{equation}
\frac{\Snull(\omega;\tilde a,\tilde \sigma^2)}{\Spop(\omega)}
=
\left(1+\frac{\delta \sigma^2}{\sigma_\epsilon^2}\right)
\frac{\Pfun_a(\omega)}{\Pfun_{a+\delta a}(\omega)}
=
1+u\tilde e_1+v\tilde e_2+O(u^2+v^2+uv),
\label{eqS:null-expand}
\end{equation}
with
\begin{equation}
u=\frac{\delta \sigma^2}{\sigma_\epsilon^2},
\qquad
v=\delta a,
\qquad
\tilde e_1(\omega)=1,
\qquad
\tilde e_2(\omega)=\frac{2(\cos\omega-a)}{\Pfun_a(\omega)}.
\label{eqS:basis}
\end{equation}
The tangent space is therefore $\mathcal{T}=\mathrm{span}\{\tilde e_1,\tilde e_2\}$.

\subsection{Population Whittle geometry}

For two spectra $S_1,S_2$, the normalized population Whittle/Kullback-Leibler divergence is
\begin{equation}
\DKL(S_1\|S_2)
=
\frac{1}{4\pi}
\int_{-\pi}^{\pi}
\left[
\frac{S_1(\omega)}{S_2(\omega)}
-\log\frac{S_1(\omega)}{S_2(\omega)}-1
\right]d\omega.
\label{eqS:DKL}
\end{equation}
Using the scalar expansion $(1+\delta)-\log(1+\delta)-1=\delta^2/2+O(\delta^3)$ gives the local quadratic reduction
\begin{equation}
\DKL\!\left(
\Strue(\cdot;\lambda)\,\|\,\Snull(\cdot;\tilde a,\tilde \sigma^2)
\right)
=
\frac{1}{4}
\bigl\|
\lambda^2 h-u\tilde e_1-v\tilde e_2
\bigr\|_{L^2}^2
+O\!\left((|u|+|v|+\lambda^2)^3\right).
\label{eqS:local-reduction}
\end{equation}

\subsection{Orthogonality of the tangent basis}

The orthogonality is exact, not heuristic. Jensen's formula gives
\begin{equation}
\frac{1}{2\pi}\int_{-\pi}^{\pi}\log \Pfun_a(\omega)\,d\omega=0
\qquad (|a|<1),
\label{eqS:Jensen}
\end{equation}
and differentiation with respect to $a$ yields
\begin{equation}
0
=
\frac{1}{2\pi}\int_{-\pi}^{\pi}
\frac{2(a-\cos\omega)}{\Pfun_a(\omega)}\,d\omega
=
-\langle \tilde e_1,\tilde e_2\rangle.
\label{eqS:orthogonality}
\end{equation}
Differentiating once more gives
\begin{equation}
\partial_a\tilde e_2
=
-\frac{2}{\Pfun_a}
+
\frac{4(\cos\omega-a)^2}{\Pfun_a^2}
=
-\frac{2}{\Pfun_a}+\tilde e_2^2,
\label{eqS:de2}
\end{equation}
from which the norm of $\tilde e_2$ follows. The tangent basis is therefore orthogonal:
\begin{equation}
\langle \tilde e_1,\tilde e_2\rangle=0,
\qquad
\|\tilde e_1\|^2=1,
\qquad
\|\tilde e_2\|^2=\frac{2}{1-a^2},
\label{eqS:gram}
\end{equation}
so the Gram determinant is positive and the local minimizer is unique.

\section{Projection Coefficients and Pseudo-True Shifts}

\subsection{Basic integrals}

For $|c|<1$,
\begin{equation}
\left\langle 1,\frac{1}{\Pfun_c}\right\rangle
=
\frac{1}{1-c^2},
\qquad
\left\langle \frac{1}{\Pfun_c},\frac{1}{\Pfun_c}\right\rangle
=
\frac{1+c^2}{(1-c^2)^3}.
\label{eqS:basic-integrals}
\end{equation}
Therefore
\begin{equation}
\|h\|^2
=
\frac{\sigma_\eta^4}{\sigma_\epsilon^4}
\frac{1+b^2}{(1-b^2)^3}.
\label{eqS:hnorm}
\end{equation}

\subsection{Projection coefficients}

The basic projection integrals are
\begin{equation}
\langle h,\tilde e_1\rangle
=
\frac{\sigma_\eta^2}{\sigma_\epsilon^2}\frac{1}{1-b^2},
\qquad
\langle h,\tilde e_2\rangle
=
\frac{\sigma_\eta^2}{\sigma_\epsilon^2}
\frac{2b}{(1-ab)(1-b^2)}.
\label{eqS:proj}
\end{equation}
The second coefficient comes from a contour integral on the unit circle whose poles lie at $z=a$ and $z=b$. The exact residue sum collapses to the rational form in Eq.~\eqref{eqS:proj}.

\subsection{Local pseudo-true shifts}

They imply the local pseudo-true shifts
\begin{equation}
\tilde \sigma^{2*}(\lambda)
=
\sigma_\epsilon^2
+
\lambda^2\frac{\sigma_\eta^2}{1-b^2}
+O(\lambda^4),
\label{eqS:sigma-star}
\end{equation}
and
\begin{equation}
\tilde a^*(\lambda)
=
a
+
\lambda^2
\frac{\sigma_\eta^2}{\sigma_\epsilon^2}
\frac{b(1-a^2)}{(1-ab)(1-b^2)}
+O(\lambda^4).
\label{eqS:a-star}
\end{equation}
These shifts are auxiliary to the main theorem but provide a direct check that the best null model moves along the one-pole manifold exactly as predicted by the projection geometry.

\begin{figure}[t]
  \centering
  \includegraphics[width=0.495\textwidth]{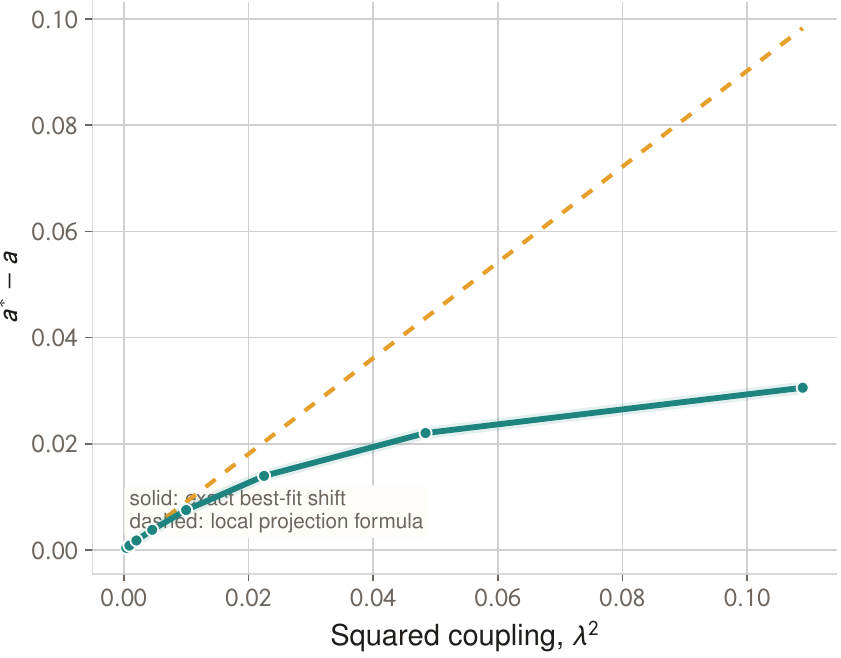}\hfill
  \includegraphics[width=0.495\textwidth]{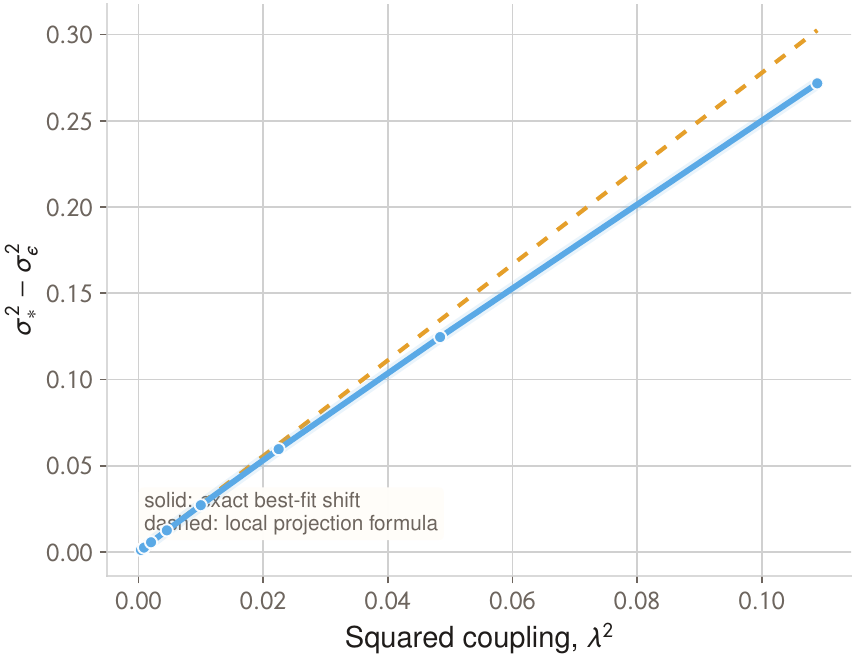}
  \caption{Auxiliary pseudo-true checks. Left: the leading-order prediction for the best-fit AR pole shift. Right: the corresponding prediction for the best-fit innovation variance shift. Both formulas track the numerical pseudo-true motion of the null model, supporting the projection picture without being part of the principal main-text claim.}
  \label{figS:pseudotrue}
\end{figure}

\section{Proof of the Quartic Detectability Theorem}

\subsection{Statement proved here}

In this appendix we prove Theorem~\ref{thm:main}. Equivalently, for $|a|<1$ and $|b|<1$, there exists $\lambda_0(a,b)>0$ such that for $|\lambda|<\lambda_0(a,b)$ the local minimizer branch exists uniquely and
\begin{equation}
\Dloc(\lambda)
=
C(a,b,\sigma_\epsilon,\sigma_\eta)\lambda^4+O(\lambda^6),
\label{eqS:quartic}
\end{equation}
with
\begin{equation}
C(a,b,\sigma_\epsilon,\sigma_\eta)
=
\frac{\sigma_\eta^4}{2\sigma_\epsilon^4}
\frac{b^2(a-b)^2}{(1-b^2)^3(1-ab)^2}.
\label{eqS:C}
\end{equation}

\begin{remark}
The theorem is pointwise in parameter space. Near coalescence, what degenerates is the quartic coefficient $C$, not the existence of the local expansion itself. The theorem controls the nearby pseudo-true branch, not the global infimum over the full stationary one-pole family.
\end{remark}

\subsection{Existence and uniqueness of the local branch}

The local quadratic reduction in Eq.~\eqref{eqS:local-reduction}, together with the positive Gram determinant from Eq.~\eqref{eqS:gram}, implies that the Hessian of the reduced objective is strictly positive definite at the null point. The implicit-function theorem therefore gives a unique local minimizer branch $(\tilde a^*(\lambda),\tilde \sigma^{2*}(\lambda))$ for sufficiently small $|\lambda|$.

\subsection{Projection formula and residual norm}

Let
\begin{equation}
\Rres=h-\Pi_{\mathcal{T}}h.
\label{eqS:Rdef}
\end{equation}
Because the tangent directions are orthogonal, the residual norm is
\begin{equation}
\|\Rres\|^2
=
\|h\|^2
-
\frac{\langle h,\tilde e_1\rangle^2}{\|\tilde e_1\|^2}
-
\frac{\langle h,\tilde e_2\rangle^2}{\|\tilde e_2\|^2}.
\label{eqS:Rraw}
\end{equation}
Substituting the exact projection integrals yields
\begin{equation}
\|\Rres\|^2
=
\frac{\sigma_\eta^4}{\sigma_\epsilon^4}
\frac{2b^2(a-b)^2}{(1-b^2)^3(1-ab)^2}.
\label{eqS:Rclosed}
\end{equation}
Together with Eq.~\eqref{eq:projection-law}, this proves Eq.~\eqref{eqS:quartic}. Positivity is immediate because every factor in Eq.~\eqref{eqS:C} is nonnegative on $|a|<1$, $|b|<1$, and the only zeros are $b=0$ or $a=b$.

\subsection{What this appendix does and does not prove}

The theorem above identifies the nearby pseudo-true branch and the asymptotic law for its local minimum. It does not prove that the same expansion controls the global infimum over the full stationary one-pole class. That stronger statement is instead checked numerically in Appendix I by exact global population minimization over the full parameter domain.

\section{Corollaries and the Local Tangent-Absorption Criterion}

\subsection{Coalescence law}

The quartic coefficient factorizes as
\begin{equation}
C(a,b,\sigma_\epsilon,\sigma_\eta)
=
\frac{\sigma_\eta^4}{2\sigma_\epsilon^4}
\frac{b^2}{(1-b^2)^3(1-ab)^2}(a-b)^2,
\label{eqS:Cfactor}
\end{equation}
so
\begin{equation}
C\propto (a-b)^2
\qquad
\text{as}\quad b\to a.
\label{eqS:coal}
\end{equation}

\subsection{Population boundary}

The corresponding leading-order population boundary is
\begin{equation}
\lcpop(N)
\!=\!
\left[
\frac{\Delta k\,\sigma_\epsilon^4(1-b^2)^3(1-ab)^2}
{\sigma_\eta^4 b^2(a-b)^2}
\frac{\log N}{N}
\right]^{1/4}
\left(1+o(1)\right),
\label{eqS:lcpop}
\end{equation}
which reduces to Eq.~\eqref{eq:lambda-c} for the effective two-pole alternative with $\Delta k=2$.

\subsection{Abstract geometric criterion}

The solvable model above isolates a broader local criterion. Let $\mathcal{M}$ be a smooth spectral manifold through $S_0$. If the leading perturbation enters as
\begin{equation}
S_{\mathrm{true}}(\lambda)=S_0(1+\lambda^2 h+O(\lambda^4)),
\label{eqS:criterion-true}
\end{equation}
and if the local Whittle Hessian on the tangent space $\mathcal{T}$ of $\mathcal{M}$ is positive definite, then the same quadratic reduction gives a unique local minimizer branch and
\begin{equation}
\Dloc(\lambda)
=
\frac{\lambda^4}{4}\|(I-\Pi_{\mathcal{T}})h\|^2+O(\lambda^6).
\label{eqS:abstract}
\end{equation}
This statement is intended as a geometric criterion, not as a stand-alone theorem for arbitrary model classes: making it theorem-grade would require explicit assumptions on parameterization, norms, and remainder control beyond what is needed for the solvable one-pole case.

\subsection{What is not proved here}

The theorem package establishes the quartic law and its exact prefactor within the one-pole projection class. It does not prove the first nonzero post-quartic coefficient at strict coalescence, and it does not claim a universal law for arbitrary hidden-variable architectures.

\section{Enriched Null Families and the ARMA$(1,1)$ Corollary}

\subsection{ARMA$(1,1)$ one-pole null family}

The simplest enriched null class keeps a single relaxation pole but adds one moving-average zero:
\begin{equation}
\Snull^{(1,1)}(\omega;\tilde a,\theta,\tilde \sigma^2)
=
\tilde \sigma^2\frac{\Pfun_\theta(\omega)}{\Pfun_{\tilde a}(\omega)}.
\label{eqS:arma-null}
\end{equation}
At the AR$(1)$ null point $(\tilde a,\theta,\tilde \sigma^2)=(a,0,\sigma_\epsilon^2)$, the relative expansion becomes
\begin{equation}
\frac{\Snull^{(1,1)}(\omega;\tilde a,\theta,\tilde \sigma^2)}{\Spop(\omega)}
=
1+u\tilde e_1+v\tilde e_2+w\tilde e_3+O(u^2+v^2+w^2+uv+uw+vw),
\label{eqS:arma-expand}
\end{equation}
with
\begin{equation}
w=\theta,
\qquad
\tilde e_3(\omega)=-2\cos\omega,
\label{eqS:e3}
\end{equation}
and the previously defined $\tilde e_1,\tilde e_2$. Thus the tangent space is enlarged to
\begin{equation}
\mathcal{T}_{(1,1)}
=
\mathrm{span}\{\tilde e_1,\tilde e_2,\tilde e_3\}.
\label{eqS:arma-tangent}
\end{equation}

\subsection{Gram matrix and projection data}

The extra tangent direction has the exact inner products
\begin{equation}
\langle \tilde e_1,\tilde e_3\rangle=0,
\qquad
\langle \tilde e_2,\tilde e_3\rangle=-2,
\qquad
\|\tilde e_3\|^2=2.
\label{eqS:arma-gram-entries}
\end{equation}
Hence, for generic $a\neq 0$, the three-dimensional Gram matrix is
\begin{equation}
G_{(1,1)}
=
\begin{pmatrix}
1 & 0 & 0 \\
0 & \dfrac{2}{1-a^2} & -2 \\
0 & -2 & 2
\end{pmatrix},
\qquad
\det G_{(1,1)}=\frac{4a^2}{1-a^2}>0.
\label{eqS:arma-gram}
\end{equation}
The nongeneric point $a=0$ is degenerate because $\tilde e_2$ and $\tilde e_3$ are collinear there ($\tilde e_2=-\tilde e_3$).

The new projection coefficient is also explicit:
\begin{equation}
\langle h,\tilde e_3\rangle
=
-\frac{2b\,\sigma_\eta^2}{\sigma_\epsilon^2(1-b^2)}.
\label{eqS:arma-proj}
\end{equation}
Together with Eq.~\eqref{eqS:proj}, this closes the three-dimensional projection problem completely.

\subsection{Closed quartic coefficient}

\begin{proposition}
\label{propS:arma}
For generic $a\neq 0$ with $|a|<1$ and $|b|<1$, local Whittle minimization over the ARMA$(1,1)$ one-pole null family gives
\begin{equation}
\Dlocarma(\lambda)
=
C_{(1,1)}(a,b,\sigma_\epsilon,\sigma_\eta)\lambda^4+O(\lambda^6),
\label{eqS:arma-law}
\end{equation}
with the exact coefficient
\begin{equation}
C_{(1,1)}
=
\frac{\sigma_\eta^4}{2\sigma_\epsilon^4}
\frac{b^4(a-b)^2}{(1-b^2)^3(1-ab)^2}
=
b^2 C.
\label{eqS:arma-C}
\end{equation}
\end{proposition}

The proof is the same projection argument as in Appendix E, but now with $\Pi_{\mathcal{T}_{(1,1)}}$ in place of $\Pi_{\mathcal{T}}$. Since the enlarged tangent space absorbs one extra component of $h$, the quartic coefficient is reduced by a factor of $b^2$ relative to the strict AR$(1)$ null, but the coalescence factor $(a-b)^2$ and the quartic onset in $\lambda$ survive unchanged. Direct numerical evaluation of the three-dimensional projection reproduces Eq.~\eqref{eqS:arma-C} to machine precision for representative parameter sets, and Appendix J reports an operational Whittle-BIC stress test with this enriched null.

\section{Independent Symbolic Verification}

\subsection{Symbolic identities checked exactly}

A separate symbolic pipeline verifies, without numerical substitution:
\begin{enumerate}
  \item the contour-integral identity that gives Eq.~\eqref{eqS:proj};
  \item the exact residual norm in Eq.~\eqref{eqS:Rclosed};
  \item the factorization of the coalescence term $(a-b)^2$;
  \item the positivity of the Gram determinant from Eq.~\eqref{eqS:gram}.
\end{enumerate}
These checks close the main algebraic loophole in the theorem package: the quartic coefficient does not rely on a single hand derivation.

\subsection{Why this matters}

The symbolic layer makes the theorem harder to attack. The coefficient no longer rests on a single contour-integral route; it is fixed both by direct analysis and by an independent algebraic verification chain.

\section{Exact Numerical KL Validation}

\subsection{Baseline parameter set}

Unless otherwise stated, the numerical tests use the baseline parameters
\begin{equation}
(a,b,\sigma_\epsilon^2,\sigma_\eta^2)=(0.95,0.8,1,1).
\label{eqS:baseline}
\end{equation}
The exact-population validation never inserts the local asymptotic formula into the objective. Instead, for each coupling value we minimize the full Whittle/KL divergence over the stationary one-pole parameter domain and then compare the resulting minimum to the theorem-level prediction.

\subsection{Quartic and threshold laws}

Exact population minimization of the Whittle objective over the one-pole class gives a quartic slope of $2.67918$ for $\lambda\le 0.15$. For the baseline parameters in Eq.~\eqref{eqS:baseline}, the quartic approximation remains accurate to within about $10\%$ for $\lambda\lesssim 0.15$. Solving $N\DKL^{\min,\mathrm{num}}(\lambda_c)=\log N$ over $N\in\{200,400,800,1600,3200,6400,12800\}$ gives a threshold slope of $-0.254360$.

\subsection{Coalescence law}

Extracting the quartic coefficient on the window $\lambda\in\{0.005,0.0075,0.01,0.015\}$ and sweeping $b$ toward $a$ gives a normalized coalescence slope of $1.922526$. At strict coalescence, $\DKL^{\min,\mathrm{num}}/\lambda^4$ keeps decreasing as $\lambda\to 0$, confirming that the quartic coefficient truly vanishes.

\subsection{Pseudo-true shifts}

The same exact numerical optimization returns the best-fit one-pole parameters. Their motion agrees with Eqs.~\eqref{eqS:sigma-star} and \eqref{eqS:a-star}, providing an additional numerical check of the projection geometry.

\begin{figure}[t]
  \centering
  \includegraphics[width=0.495\textwidth]{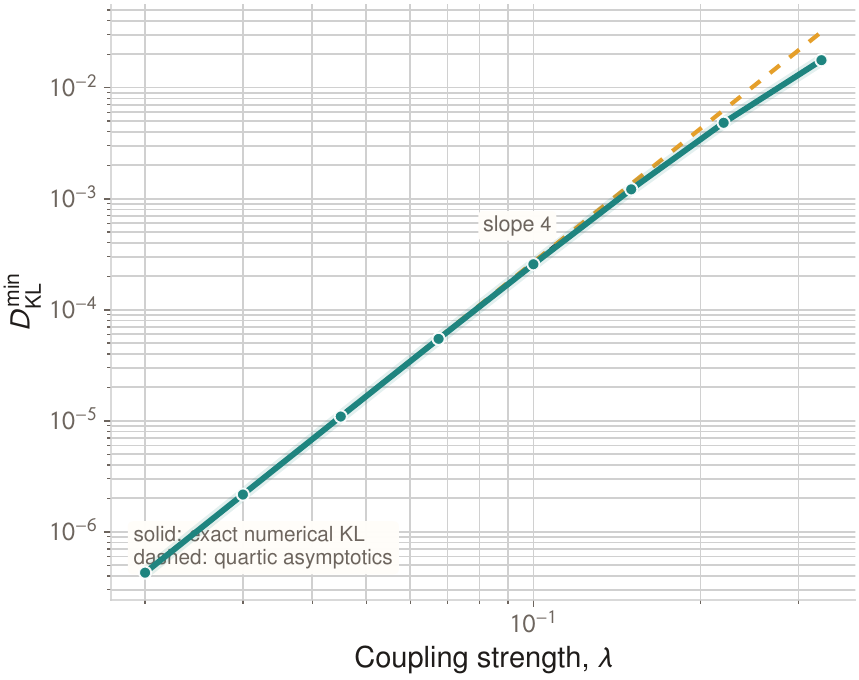}\hfill
  \includegraphics[width=0.495\textwidth]{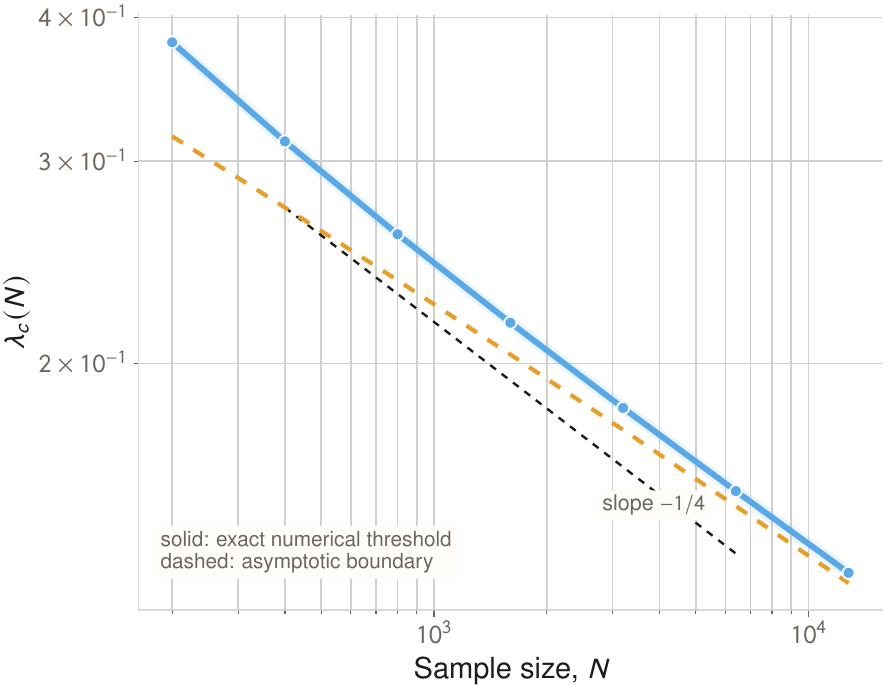}
  \caption{Supplementary numerical validation. Left: exact numerical Whittle minimization confirms the quartic coupling law. Right: the same population solver confirms the threshold scaling predicted by the closed-form coefficient.}
  \label{figS:validation}
\end{figure}

\begin{figure}[t]
  \centering
  \includegraphics[width=0.495\textwidth]{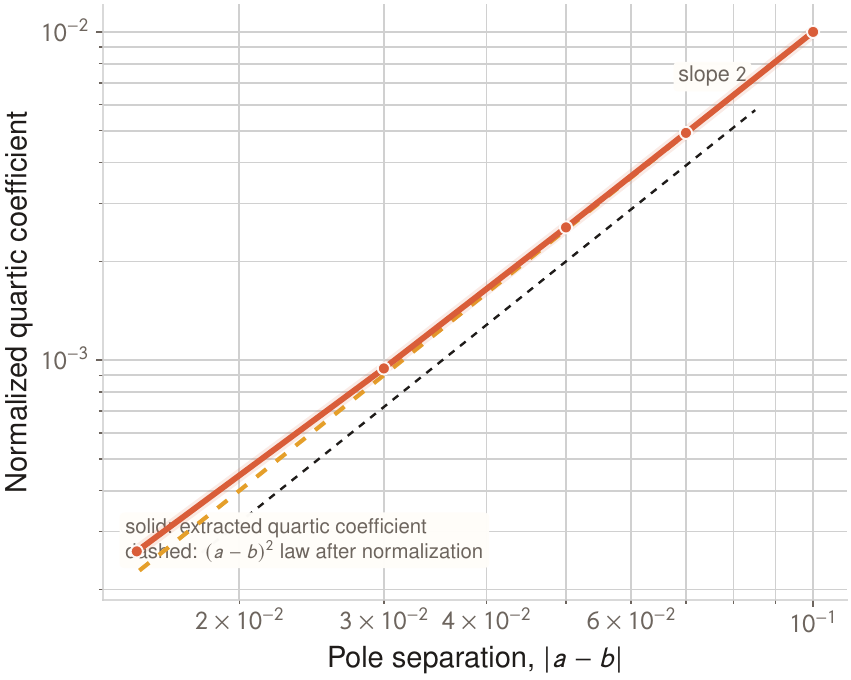}\hfill
  \includegraphics[width=0.495\textwidth]{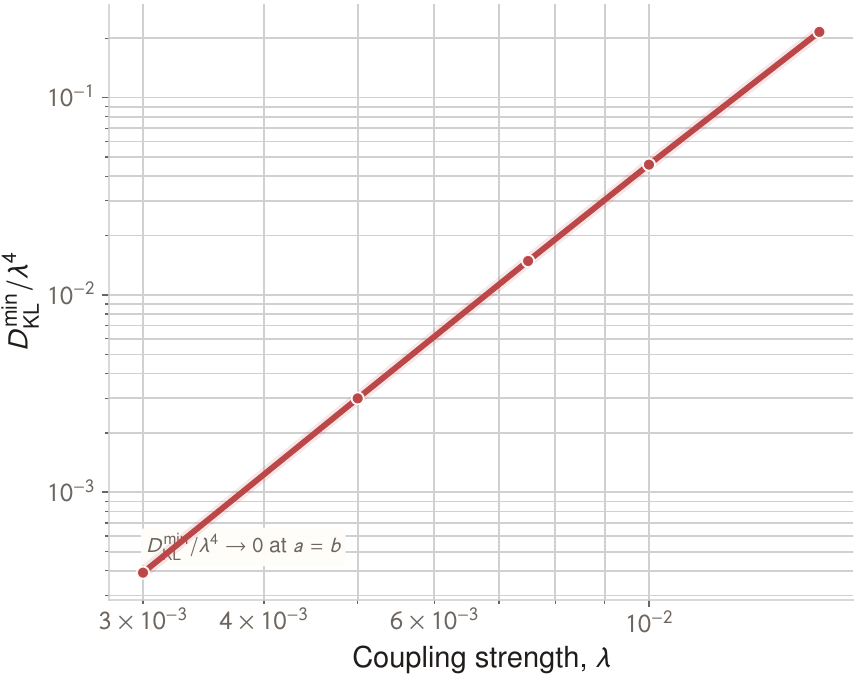}
  \caption{Coalescence validation. Left: the extracted quartic coefficient collapses as $(a-b)^2$. Right: at strict coalescence, $\Dloc(\lambda)/\lambda^4$ keeps decreasing as $\lambda\to 0$, confirming that the quartic coefficient vanishes.}
  \label{figS:coalescence}
\end{figure}

\section{Operational Monte Carlo Validation}

\subsection{Model-selection rule}

The operational model-selection benchmark compares the one-pole null against the effective two-pole alternative
\begin{equation}
S_{\mathrm{alt}}(\omega)
\!=\!
\frac{\sigma^2}{\Pfun_{\tilde a}(\omega)}
+
\frac{q}{\Pfun_{\tilde a}(\omega)\Pfun_{\tilde b}(\omega)},
\label{eqS:alt}
\end{equation}
with $\Delta k=2$. We fit both models by the Whittle likelihood and declare the hidden driver detected when ${\rm BIC}_{\rm alt}<{\rm BIC}_{\rm null}$. Only the observed series $\{X_t\}_{t=1}^N$ is retained in each synthetic trial. This benchmark is intentionally favorable in one respect: the alternative contains the exact observed spectrum family generated by Eq.~\eqref{eq:model}. Its role is therefore not to prove robustness to arbitrary model uncertainty, but to ask whether the theorem-scale boundary is already visible when the alternative structure is known. All reported intervals are Wilson 95\% intervals for binomial detection frequencies.

\subsection{Collapse experiment}

For the strengthened experiment, we use $200$ repetitions per grid point, sample sizes $N=256,512,1024,2048$, and scaled couplings $\lambda/\lcpop(N)\in\{0.6,0.8,1.0,1.2,1.5\}$. At the predicted boundary, the detection probabilities are $0.385$, $0.410$, $0.535$, and $0.505$. The corresponding 95\% Wilson intervals are approximately $[0.320,0.454]$, $[0.344,0.479]$, $[0.466,0.603]$, and $[0.436,0.574]$. The mean spread across $N$ at fixed scaled coupling is $0.138$, and the mean Wilson width at the predicted boundary is about $0.136$. The interpolated $50\%$-power points are $r_{50}(N)=1.100$, $1.053$, $0.982$, and $0.997$.

\subsection{Role in the main-text package}

This experiment is the bridge from theorem to operational relevance. The strongest defensible claim is not exact finite-size collapse, but a common turnover near the predicted scaled boundary together with convergence of the effective crossover toward unity.

\subsection{Hard synthetic stress tests}

\begin{figure}[t]
  \centering
  \includegraphics[width=0.495\textwidth]{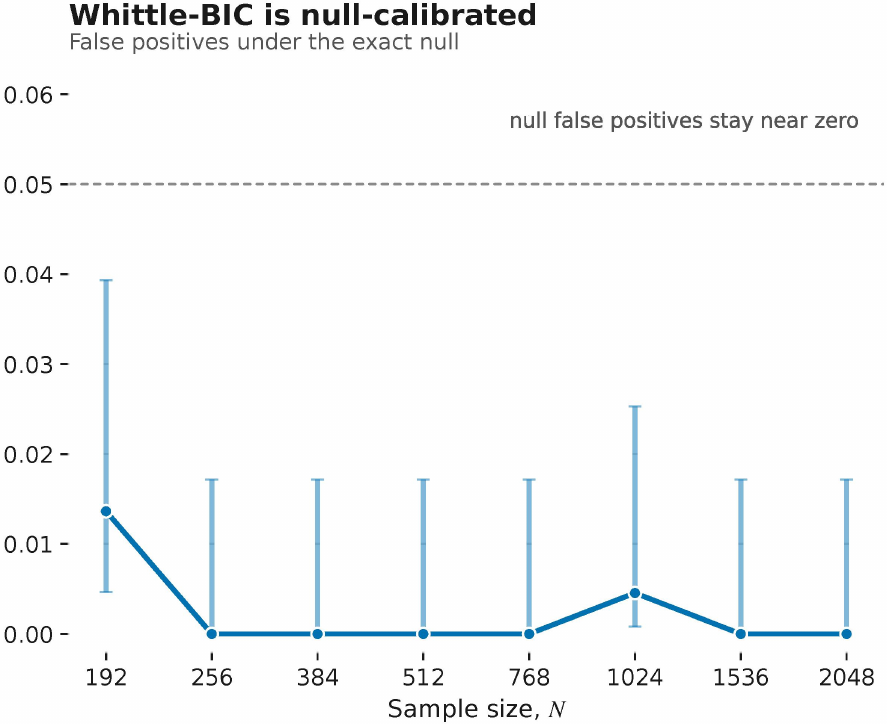}\hfill
  \includegraphics[width=0.495\textwidth]{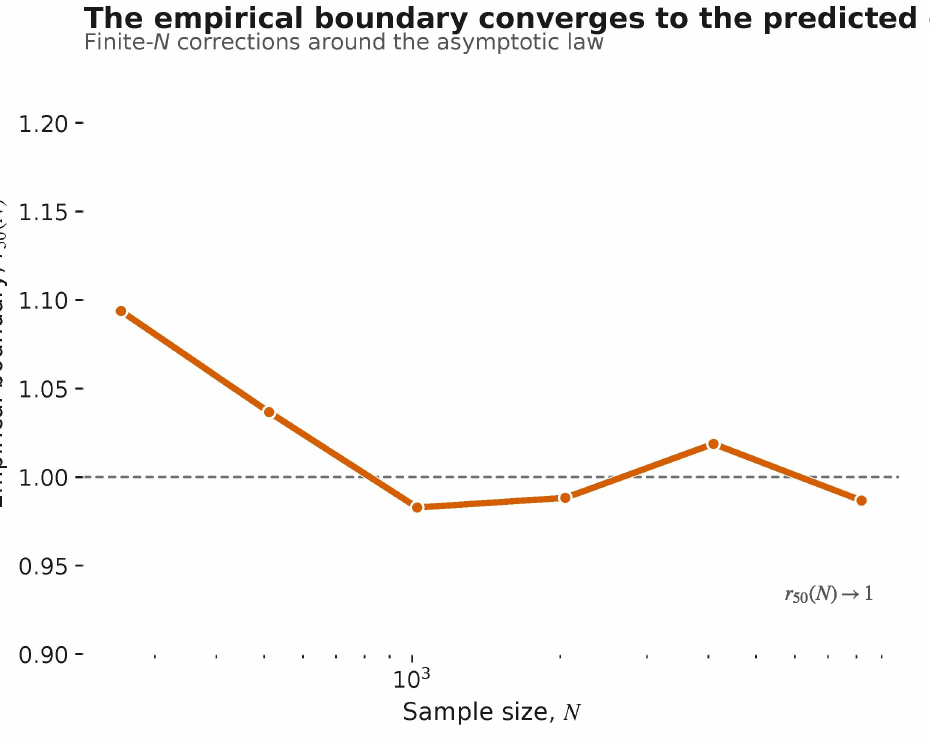}\\[0.8em]
  \includegraphics[width=0.495\textwidth]{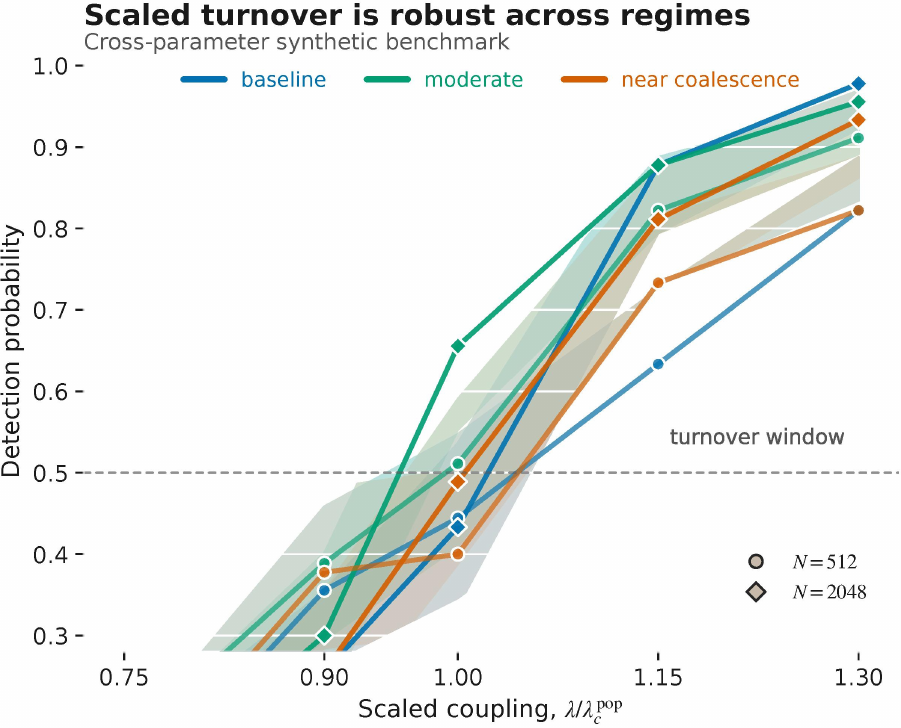}\hfill
  \includegraphics[width=0.495\textwidth]{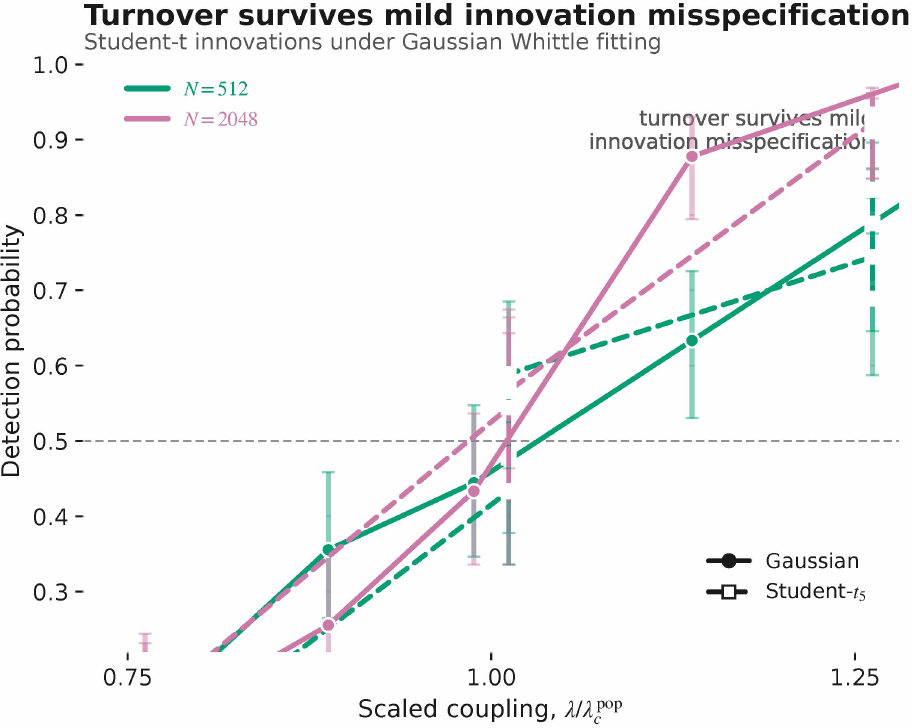}
  \caption{Hard synthetic stress tests. Top left: the operational Whittle-BIC procedure is null-calibrated. Top right: the empirical boundary converges toward the predicted scaled boundary. Bottom left: the turnover picture persists across distinct $(a,b)$ regimes. Bottom right: it survives mild innovation misspecification.}
  \label{figS:stress}
\end{figure}

Null calibration remains stringent: under $\lambda=0$, the false-positive rate stays at or below $0.5\%$ for $N=256$ through $2048$, with Wilson upper bounds below $0.026$ and mean $\Delta{\rm BIC}$ values between about $-8.8$ and $-13.3$ in favor of the null. The same scaled-turnover picture also persists across well-separated, moderate, and near-coalescent parameter regimes, as shown in Fig.~\ref{figS:stress}. For these three regimes, the detection power at $\lambda/\lcpop=1.0$ spans the range $0.400$ to $0.656$, while at $\lambda/\lcpop=1.15$ it spans $0.633$ to $0.878$. Near coalescence, the mean $\Delta{\rm BIC}$ is slightly negative at the nominal boundary for smaller $N$, but becomes strongly positive above it, exactly as the asymptotic interpretation predicts.

\subsection{Mild innovation misspecification}

The operational boundary is not restricted to exact Gaussian innovations. Keeping the same linear dynamics but replacing both innovations by standardized Student-$t_5$ noise leaves the boundary-region picture intact. For the baseline parameters, the Whittle-BIC detection powers are $0.433$ and $0.544$ at $\lambda/\lcpop=1.0$ for $N=512$ and $2048$, with corresponding Wilson intervals $[0.336,0.536]$ and $[0.442,0.643]$. At $\lambda/\lcpop=1.25$, the powers rise to $0.833$ and $0.922$, with Wilson intervals $[0.743,0.896]$ and $[0.848,0.962]$. This test does not preserve every exact theorem-level coefficient, but it does show that turnover near the predicted scaled boundary survives a mild distributional mismatch between generator and fitted model.

\subsection{Enriched-null operational stress test}

A more demanding null replaces the strict AR$(1)$ fit by the enriched one-pole ARMA$(1,1)$ family from Appendix G. The null then has one extra parameter, so the model-count gap drops to $\Delta k=1$ and the relevant scaled coupling is the corresponding local boundary $\lambda_{c,(1,1)}^{\mathrm{pop}}$. For the baseline parameters and $80$ repetitions per grid point, the detection powers at $\lambda/\lambda_{c,(1,1)}^{\mathrm{pop}}=\{0.8,1.0,1.15\}$ are $(0.188,0.412,0.650)$ for $N=512$ and $(0.250,0.563,0.775)$ for $N=2048$. The interpolated $50\%$-power locations are $r_{50}(512)\approx 1.06$ and $r_{50}(2048)\approx 0.97$. Thus the enriched null weakens detectability substantially, exactly as the analytic coefficient $C_{(1,1)}=b^2 C$ predicts, but it does not destroy the boundary-centered turnover. The runs are generated by the dedicated ARMA-null stress script and summarized alongside the main stress suite in the repository results.

\begin{figure}[t]
  \centering
  \includegraphics[width=0.68\textwidth]{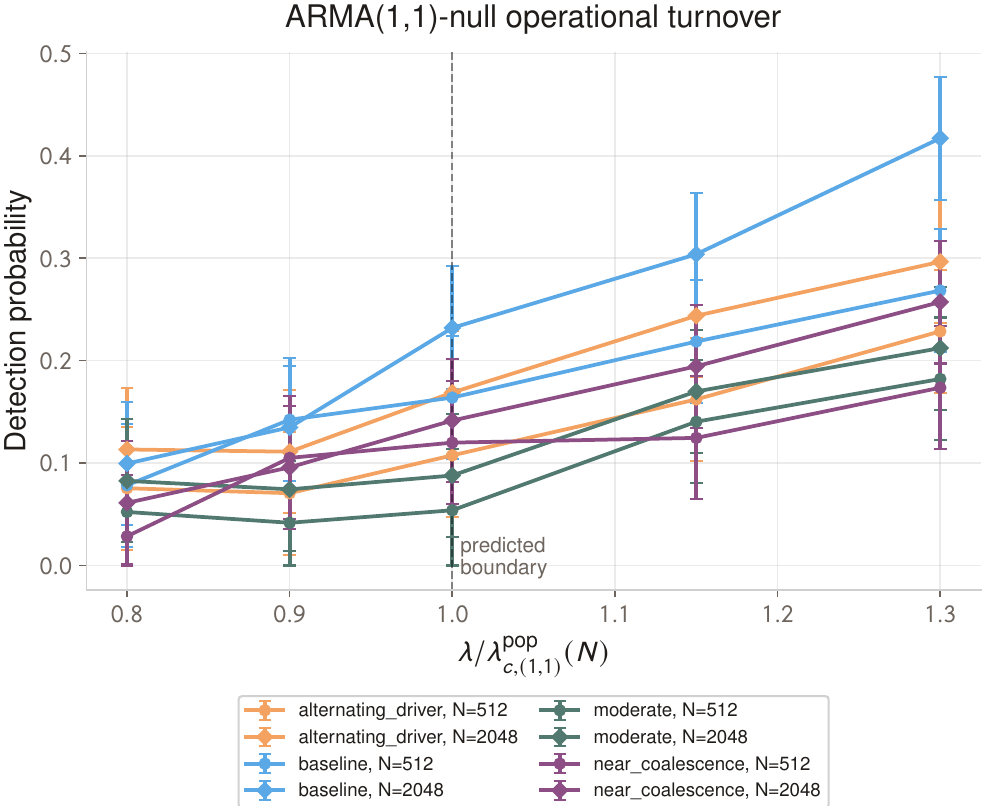}
  \caption{Operational turnover under an enriched ARMA$(1,1)$ null. The strict AR$(1)$ null is replaced by the one-pole ARMA$(1,1)$ family, and detection probability is plotted against the corresponding local boundary $\lambda_{c,(1,1)}^{\mathrm{pop}}(N)$ for $N=512$ and $2048$. The turnover persists, but the enriched null shifts the crossover upward at smaller $N$, consistent with the reduced quartic coefficient $C_{(1,1)}=b^2 C$.}
  \label{figS:arma-null}
\end{figure}

\section{Continuous-Time Foundation and Scope}

\subsection{Continuous-time driven OU foundation}

The discrete-time theorem was motivated by the continuous-time driven Ornstein-Uhlenbeck system
\begin{equation}
\dot X=-\gamma X+\lambda F+\sqrt{2D}\,\xi_X(t),
\qquad
\dot F=-\alpha F+\sqrt{2D_F}\,\xi_F(t).
\label{eqS:OU}
\end{equation}
Its exact observed spectrum is
\begin{equation}
S_X(\omega)
\!=\!
\frac{2D}{\gamma^2+\omega^2}
+
\frac{2D_F\lambda^2}{(\gamma^2+\omega^2)(\alpha^2+\omega^2)}.
\label{eqS:OU-spectrum}
\end{equation}
When $\alpha=\gamma$, the leading perturbation is proportional to $(\gamma^2+\omega^2)^{-2}$ and becomes tangent to the one-pole OU manifold. This is the continuous-time precursor of the discrete-time coalescence law.

\subsection{Why the theorem is completed in discrete time}

The continuous-time picture is structurally correct but less convenient for a complete theorem because the frequency domain is unbounded and the weighted absolute-space formulation requires additional care at high frequency. In discrete time, the frequency interval is compact and the projection algebra closes cleanly in standard $L^2$.

\subsection{Back-of-the-envelope physical scale estimates}

The climate-style relevance statement in the main text can be quantified directly from Eq.~\eqref{eqS:lcpop}. For equal noise scales and $(a,b)=(0.90,0.80)$, the population boundary is
\begin{equation}
\lcpop(10^3)\approx 0.30,
\qquad
\lcpop(10^4)\approx 0.18.
\label{eqS:phys-baseline}
\end{equation}
For the near-coalescent choice $(a,b)=(0.90,0.88)$, the same formula gives
\begin{equation}
\lcpop(10^3)\approx 0.39,
\qquad
\lcpop(10^4)\approx 0.23.
\label{eqS:phys-near}
\end{equation}
Thus modest pole coalescence already raises the leading-order threshold by a visible amount. More generally, at fixed scaled coupling $\lambda/\lcpop$,
\begin{equation}
\frac{N}{\log N}\propto |a-b|^{-2},
\label{eqS:data-cost}
\end{equation}
so shrinking the pole separation from $0.10$ to $0.02$ raises the leading-order data requirement by roughly a factor of $25$. These are not application-specific claims; they are order-of-magnitude consequences of the exact solvable boundary law.

\subsection{Scope and non-claims}

The present theorem package proves:
\begin{itemize}
  \item an exact quartic detectability law in the one-pole projection class;
  \item an exact closed form for the quartic coefficient;
  \item a quantitative coalescence law $C\propto(a-b)^2$;
  \item a leading-order population boundary $\lcpop\propto (\log N/N)^{1/4}$;
  \item symbolic, numerical, and operational support for these statements.
\end{itemize}
It does not claim a universal theorem for arbitrary hidden-variable architectures, nor the first nonzero post-quartic coefficient exactly at strict coalescence. For richer null families, one expects a larger tangent space, a smaller leading normal component, and therefore a weaker detectability coefficient; if the null family can already represent the driver-induced multipole structure, the first nonzero detectability term may occur at still higher order.

\section{Extension to an AR\texorpdfstring{$(2)$}{(2)} Hidden Driver}
\label{app:ar2}

\subsection{Model and spectrum}

The AR$(2)$-driven extension replaces the hidden AR$(1)$ process by
\begin{equation}
F_{t+1}=b_1 F_t+b_2 F_{t-1}+\eta_t,
\label{eqS:ar2-driver}
\end{equation}
with characteristic roots $z_1,z_2$ satisfying $z^2-b_1 z-b_2=0$. Stationarity requires both roots to lie inside the unit disk; the stability region is the triangle $b_1+b_2<1$, $b_2-b_1<1$, $b_2>-1$. The AR$(2)$ spectral polynomial is
\begin{equation}
Q_b(\omega)=|1-b_1 e^{-i\omega}-b_2 e^{-2i\omega}|^2=\prod_{j=1}^{2}\Pfun_{z_j}(\omega)
\label{eqS:Q-def}
\end{equation}
when the roots are real (overdamped regime), where the last equality uses the factorization through $P_{z_j}(\omega)=|e^{i\omega}-z_j|^2$. For complex-conjugate roots (oscillatory regime), $Q_b$ remains real and positive but does not decompose into real one-pole factors.

The observed spectrum of $X$ is
\begin{equation}
\Strue(\omega;\lambda)
=
\frac{\sigma_\epsilon^2}{\Pfun_a(\omega)}
+
\frac{\lambda^2\sigma_\eta^2}{\Pfun_a(\omega)\,Q_b(\omega)},
\label{eqS:ar2-spectrum}
\end{equation}
so the relative perturbation becomes $h(\omega)=\sigma_\eta^2/[\sigma_\epsilon^2\,Q_b(\omega)]$. Because the null manifold is still the one-pole AR$(1)$ family, the tangent space $\mathcal{T}=\mathrm{span}\{\tilde e_1,\tilde e_2\}$ and the Gram data are unchanged from the baseline.

\subsection{Quartic law and absence of coalescence suppression}

The abstract tangent-absorption criterion (Appendix~F, Eq.~\eqref{eqS:abstract}) applies directly:
\begin{equation}
\Dloc(\lambda)
=
\frac{\lambda^4}{4}\|\Rres\|_{L^2}^2+O(\lambda^6),
\qquad
\Rres=h-\Pi_{\mathcal{T}}h.
\label{eqS:ar2-quartic}
\end{equation}
The quartic coefficient $C_{\mathrm{AR}(2)}=\frac{1}{4}\|\Rres\|^2$ is computed from the standard projection $\|\Rres\|^2=\|h\|^2-\langle h,\tilde e_1\rangle^2-\langle h,\tilde e_2\rangle^2/\|\tilde e_2\|^2$, where the inner products involve three-pole residue sums on the unit circle.

The key new result is the absence of coalescence suppression. For the baseline AR$(1)$ driver, $C=0$ exactly when $b=a$ because the perturbation $h\propto 1/\Pfun_b$ becomes proportional to the null spectrum shape and is fully absorbed by the tangent space. For the AR$(2)$ driver, the perturbation $h\propto 1/Q_b$ retains structure from the second root even when one root matches $a$:
\begin{equation}
C_{\mathrm{AR}(2)}>0
\qquad
\text{for all nonwhite stationary AR$(2)$ drivers},
\label{eqS:ar2-no-coal}
\end{equation}
including the special case $z_1=a$. The richer spectral shape of the AR$(2)$ driver cannot be reabsorbed by a two-parameter reparametrization of the one-pole null. The spectrally dark regime is therefore specific to the simplest AR$(1)$ hidden dynamics and does not survive the introduction of richer driver structure.

For oscillatory drivers (complex-conjugate roots with modulus $r=\sqrt{-b_2}$), $C_{\mathrm{AR}(2)}>0$ for all real $a$, reinforcing the conclusion that oscillatory hidden dynamics cannot become spectrally dark through timescale matching.

\subsection{Numerical validation}

Population-level minimization of the Whittle KL over the one-pole null confirms quartic onset for representative parameter sets:
\begin{itemize}
  \item Overdamped $(z_1,z_2)=(0.70,0.50)$: log-log slope $3.87$.
  \item Oscillatory moderate $|r|\approx 0.71$: slope $3.95$.
  \item Oscillatory persistent $|r|\approx 0.77$: slope $3.81$.
\end{itemize}
The deviations from $4.0$ are consistent with higher-order corrections at the tested coupling range. The threshold scaling $\lcpop(N)\propto(\log N/N)^{1/4}$ is confirmed with a fitted exponent of $-0.251$.

A root-position sweep with one root varying and $z_2=0.5$ fixed shows that $C_{\mathrm{AR}(2)}$ remains strictly positive across the entire range, including at exact single-root coalescence ($z_1=a=0.95$). Whittle-BIC Monte Carlo experiments with $200$ repetitions per grid point show detection-probability turnover near the predicted scaled boundary, confirming that the operational interpretation carries over from the baseline model.

\end{document}